\documentclass{aa}
\usepackage{times}
\usepackage{graphicx}
\begin{document}
\thesaurus{08.09.2 RX J0019.8; % stars individual
           02.01.2;            % accretion disks
           08.02.1;            % binaries: close
           08.02.2;            % binaries: eclipsing
           13.25.5             % X-rays: stars     
           }
\title{
   Detailed optical studies of the galactic supersoft X-ray
   source QR And (RX\,J0019.8\,+2156)
\thanks{ 
   Based on observations at the German-Spanish Astronomical
   Center, Calar Alto, Spain and at the Wendelstein Observatory, Germany}
      }
\author{
        B.~Deufel\inst{1,2}\and 
        H.~Barwig\inst{2}\and 
        D.~\v{S}imi\'c\inst{3}\and  
        S.~Wolf\inst{2}\and   
        N.~Drory\inst{2}
       }

\offprints{bed@mpa-garching.mpg.de}
\institute{
       Max-Planck-Institut f\"ur Astrophysik, Karl
       Schwarzschildstr.~1, D-85740 Garching, Germany
\and 
       Universit\"ats-Sternwarte M\"unchen, Scheinerstr.~1,   
       81679 M\"unchen, Germany
\and   
       Max-Planck-Institut f\"ur extraterrestrische Physik,
       Giessenbach Str.~1, D-85740 Garching, Germany}
\date{Received 7 September 1998;Accepted }
%
%_____________________________________________________________________

\maketitle
\begin{abstract}
We present high-speed long term photometric observations from 1992 to
1996 \em(UBVRI)\rm, as well as recently obtained (1997) spectrophotometric
and high resolution spectroscopic studies of the bright galactic supersoft
X-ray source RX\,J0019.8\,+2156. Our photometry reveals a highly
variable object. The shortest observed quasi-periodic variations are
humps with a period of $\approx$1.8\,h. The timings of the main minimum are
not exactly regular and occur on average around phase $\phi$\,$=$\,0.0. With
our new data set we calculated an update of the orbital period.

Our simultaneous spectroscopic and spectrophotometric studies from
1997 give an even more detailed insight: blue and red shifted
satellite lines at \ion{He}{ii} ${\lambda}$\,4686 and at the strong
Balmer lines, which are interpreted as high velocity outflows, can
clearly be detected. We also present for the first time an analysis of
this supersoft source by means of Doppler tomography.  The lines of
\ion{He}{ii} ($\lambda\lambda$\,4542,\,4686), H$_{\alpha}$ and
H$_{\beta}$ show no emission from an accretion disk. These emission
lines originate in regions within the binary system with very low
velocities. The spatial origin of the emitting material is not quite
clear and will be investigated in a follow-up paper.


\end{abstract}
%
%__________________________________________________________________

\section{Introduction}

Supersoft X-Ray sources (hereafter called SSS) were discovered by \sc
Einstein\rm, established by \sc Rosat \rm (Tr\"umper et al.  1991) and
are now commonly accepted to be close binary systems with a white
dwarf primary. Little is known about the nature of the companion
star. The accretion rate in these systems is relatively high (in the
order of $10^{-7}M_{\sun}$y$^{-1}$) so that the accreted hydrogen
burns steadily on the surface of the white dwarf (van den Heuvel et
al. 1992). Several SSS are now known, most of them in the LMC and in
M31. A detailed list can be found in Greiner (1996). Only two SSS are
known in our Galaxy; one of them is RX\,J0019.8+2156 (hereafter called
RXJ0019), which was discovered by Beuermann et al. (1995).

SSS are characterized by their high X-ray luminosity from 10$^{36}$
erg\,s$^{-1}$ to 10$^{38}$ erg\,s$^{-1}$ and by extremely soft X-ray
spectra with blackbody temperatures between 20\,--\,60 eV. Most of the
optical luminosity probably originates in the bright accretion disk:
in a radially extended and vertically elevated accretion disk rim a
substantial fraction of the soft X-ray flux from the white dwarf is
reprocessed to optical and UV-light. This model successfully accounts
for features observed in the optical lightcurves of some SSS such as
CAL\,87, RXJ0019 and RX\,J0513.9-6951 (Meyer-Hofmeister et
al. 1997). Most of the optical lightcurves presented in this paper
have also been used by Meyer-Hofmeister et al. (1998) for their
analysis . They argue that the short time variability is caused by
changes in the height of the accretion disk rim. An elevated rim is
therefore the premise for explaining the luminosity in optical
bands. This picture of a bright accretion disk is also invoked by
other authors (Popham \& Di\,Stefano 1996, Matsumoto \& Fukue 1998).

RXJ0019 is a unique object. It is the brightest and nearest among the
known SSS. With its position well outside the galactic plane and its
visual brightness of $m_V$\,$\approx$\,12.5$^m$ RXJ0019 offers an
outstanding opportunity to investigate a SSS very closely with
indirect imaging techniques such as Doppler tomography.  Here we
report on such an approach to investigate RXJ0019.

In Sect. 2 we refer to our observations of RXJ0019
and the data reduction procedures used. Section 3 deals with the results
and analysis of our data. Finally, in Sect. 4 a summary and discussion
of this paper are given.

%__________________________________________________________________

\section{Observations and data}

RXJ0019 has been monitored regularly at the Wendelstein Observatory
from 1992 to 1995 (Will
\& Barwig 1996). Follow-up observations were made in autumn 1996. We
were able to observe this object on 15 nights between October and December
1996. 
 
A special approach to investigate RXJ0019 was made in October 1997. We
performed simultaneous spectroscopy and spectrophotometry at the
3.5\,m and 2.2\,m telescope at Calar Alto, respectively. Detailed
information about the individual observing campaigns is
given in Tables \ref{photo} to \ref{twin}.

\subsection{Photometry}

High-speed photometric observations of RXJ0019 were performed using
the multichannel multicolor photometer MCCP (Barwig et al. 1987) at
the 80\,cm telescope at the Wendelstein Observatory in the Bavarian
Alps. With this photometer we can monitor the object, a nearby
comparison star and the sky background simultaneously in \em UBVRI\rm.
The MCCP allows a nearly complete elimination of atmospheric
transparency variations and extinction effects (Barwig et al. 1987),
by subtracting the sky background of each color channel and dividing
the object by the comparison star measurement afterwards. Therefore,
photometric measurements were possible even under non-photometric
conditions. During the 1996 run we were able to add another 88.7 hours
of observation time for RXJ0019 to our existing dataset.

As comparison star we used the star at $RA=0^h19^m31\fs2$,
$DEC=+21\degr53'58\farcs9$ (for epoch 2000.0).  The intensities of the
light-curves presented in Fig. \ref{lc} are calculated relative to
this star. The error in the relative count rate is $\pm0.012$. In
order to normalize the individual channels, calibration measurements
were performed during photometric conditions once or twice a
night. The constancy of the calibration coefficients is
indicative of the high stability of our detectors. All observations
were made with an integration time of 2\,sec. except on the night of
Dec 10,1996 where we used an integration time of 1\,sec. A detailed
journal of the observations is given in Table \ref{photo}.

\begin{table}
\caption{\label{photo} Journal of photometric observations with
MCCP. HJD\,=\, truncated Heliocentric Julian Date $-2440000.0$,
h\,=\,hours, IT\,=\,integration time}
\begin{tabular}{ccccc}
\hline\hline 
Date & Start [HJD] & Stop [HJD] & Duration [h] & IT [s] \\
\hline
21/09/1992 & 08887.296 & 08887.668 & 8.93  & 2 \\
22/10/1995 & 10013.225 & 10013.648 & 10.16 & 2 \\
23/10/1995 & 10014.249 & 10014.637 & 9.31  & 2 \\
\hline
11/10/1996 & 10368.350 & 10368.664 & 7.537 & 2 \\ 
%\hline 
12/10/1996 & 10369.261 & 10369.655 & 9.442 & 2 \\ 
%\hline 
13/10/1996 & 10370.442 & 10370.654 & 5.111 & 2 \\ 
%\hline 
18/10/1996 & 10375.321 & 10375.552 & 5.530 & 2 \\ 
%\hline 
22/10/1996 & 10379.482 & 10379.637 & 3.704 & 2 \\ 
%\hline
23/10/1996 & 10380.237 & 10380.476 & 5.746 & 2 \\ 
%\hline 
02/11/1996 & 10390.401 & 10390.557 & 3.630 & 2 \\ 
%\hline 
03/11/1996 & 10391.224 & 10391.614 & 9.361 & 2 \\ 
%\hline 
04/11/1996 & 10392.240 & 10392.599 & 8.597 & 2 \\
%\hline 
09/11/1996 & 10397.218 & 10397.507 & 6.943 & 2 \\ 
%\hline 
10/12/1996 & 10428.394 & 10428.499 & 2.515 & 1 \\ 
%\hline 
11/12/1996 & 10429.248 & 10429.417 & 4.406 & 2 \\ 
%\hline 
12/12/1996 & 10430.250 & 10430.445 & 4.697 & 2 \\ 
%\hline 
15/12/1996 & 10433.210 & 10433.480 & 6.478 & 2 \\ 
%\hline
16/12/1996 & 10434.276 & 10434.486 & 5.051 & 2 \\ 
\hline
\end{tabular}
\end{table}
\begin{table}
\caption{\label{meka} Journal of spectrophotometric observations with
MEKASPEK. HJD\,=\,truncated Heliocentric Julian Date $-2450000.0$,
h\,=\,hours, IT\,=\,integration time}
\begin{tabular}{ccccc}
\hline\hline
Date & Start [HJD] & Stop [HJD] & Duration [h] & IT [s] \\
\hline
27/10/1997 & 749.418 & 749.628 & 4.23 & 2 \\
%\hline
30/10/1997 & 752.295 & 752.623 & 7.74 & 2 \\
\hline
\end{tabular}
\end{table}
\begin{table}
\caption{\label{twin} Journal of spectroscopic observations. B=blue
spectral region, R=red spectral region, UT=universal time, IT=integration time}
\begin{tabular}{ccccc}
\hline\hline
Date & Spectra & Phase range & Start [UT] & IT [s] \\
\hline
           &  R &  0.12--0.15 & $22^h22^m30^s$ & 1800 \\
%\cline{2-5}
           & B R & 0.16--0.19 &  $23^h11^m55^s$ & 1800 \\
%\cline{2-5}
           & B R & 0.20--0.23 & $23^h48^m25^s$ & 1800 \\
%\cline{2-5}
27-28/10/1997  & B R & 0.24--0.27 & $00^h23^m42^s$ & 1800 \\
%\cline{2-5}
           & B R & 0.28-0.31 & $01^h00^m01^s$ & 1800 \\
%\cline{2-5}
           & B R & 0.33--0.37 & $01^h50^m13^s$ & 1800 \\
%\cline{2-5}
           & B   & 0.37--0.40 & $02^h25^m04^s$ & 1800 \\
\hline
28-29/10/1997 & B R & 0.69--0.72 & $22^h56^m57^s$ & 1800 \\
\hline
           & B R & 0.44--0.47 & $18^h54^m08^s$ & 1800 \\
%\cline{2-5}
           & B R & 0.48--0.51 & $19^h28^m49^s$ & 1800 \\
%\cline{2-5}
           & B R & 0.51--0.54 & $20^h03^m14^s$ & 1800 \\
%\cline{2-5}
           & B R & 0.55--0.58 & $20^h37^m39^s$ & 1800 \\
%\cline{2-5}
           & B R & 0.59--0.61 & $21^h21^m18^s$ &  690 \\
%\cline{2-5}
           & B R & 0.69--0.72 & $22^h52^m35^s$ & 1800 \\
%\cline{2-5}
30-31/10/1997 & B R & 0.73--0.76 & $23^h26^m55^s$ & 1800 \\
%\cline{2-5}
           & B R & 0.76--0.79 & $00^h01^m13^s$ & 1800 \\
%\cline{2-5}
           & B R & 0.80--0.83 & $00^h35^m47^s$ & 1800 \\
%\cline{2-5}
           & B R & 0.83--0.86 & $01^h10^m30^s$ & 1800 \\
%\cline{2-5}
           & B R & 0.87--0.90 & $01^h46^m23^s$ & 1800 \\
%\cline{2-5}
           & B R & 0.91--0.94 & $02^h20^m45^s$ & 1800 \\
%\cline{2-5}
           & B R & 0.95--0.98 & $02^h55^m20^s$ & 1800 \\
\hline
\end{tabular}
\end{table}

\subsection{Spectrophotometry}

We observed RXJ0019 on October 27 and 30,\,1997 with MEKASPEK attached
to the 2.2\,m telescope at the Calar Alto Observatory. MEKASPEK is a
four channel fiber-optic spectrophotometer, developed at the
Universit\"ats-Sternwarte M\"unchen. With MEKASPEK we can also perform
simultaneous measurements of the object, a nearby comparison star and
the sky background within the spectral range of 3700...9000{\AA} at a
spectral resolution of $\lambda / \Delta\lambda \approx 50$. The
photon-counting two-dimensional detector (MEPSICRON) has a time
resolution of up to 5\,ms. 

Furthermore, MEKASPEK performs a correct treatment of
atmospheric effects and allows an accurate transformation to any
broadband photometric system. For more details see Mantel et
al. (1993) and Mantel \& Barwig (1993). Again, atmospheric effects
are eliminated using the standard reduction method (Barwig et
al. 1987).

We used an integration time of 2\,sec. Each night two calibration
measurements were performed to get the calibration coefficients for
the normalization of the color channels.  We also made flat-field
measurements and a measurement of a HgArRb-lamp spectrum for
wavelength calibration at the end of each night. Due to unfortunate
weather conditions only two of four program nights could be
used. Therefore, we could not cover a complete orbital phase of
RXJ0019.  The details of the MEKASPEK observations are listed in
Table \ref{meka}.

\subsection{Spectroscopy}

Simultaneously with the spectrophotometric measurements we obtained
high-resolution spectroscopic data with the Cassegrain double beam
spectrograph (TWIN) attached to the 3.5\,m telescope at Calar
Alto. The blue and red channel were equipped with the low-noise CCDs
(SITe\#6a and SITe\#4d) with a pixel size of 15$\mu$m and a CCD size
of 800$\times$2000 pixel. We used a slit width of $1\farcs 5$. For the
blue channel we chose grating T07 with a spectral resolution of
0.81\AA /pixel and a spectral range of 3300...5000\AA\space (observations
performed in second order) and for the red channel grating T04 with a
spectral resolution of 1.08\AA /pixel in a spectral range between
5500...9000\AA. The object was trailed along the slit in order to get
phase-resolved, high resolution spectra. Exposure times ranged between
690 and 1800s and the mean trail velocity was set to 320$\arcsec$/h.
The trailed spectra were binned in the direction of the slit by a factor
of 2. Helium-Argon wavelength calibration spectra were taken
approximately every hour. During the first
observing night no calibration spectra in the blue spectral range
could be obtained due to technical problems. Flatfield measurements were
taken at the end of the last observing night. Further information about
the observation is given in Table \ref{twin}. The data reduction
included bias-subtraction, flatfield-correction, sky-subtraction,
cosmic-ray elimination and wavelength-calibration as described by
Horne (1986). 

After that, the continuum of the spectroscopic data
set was calibrated with our simultaneously obtained spectrophotometric
data. By applying this method we were able to correct for intensity
variations of the emission lines which might have been caused by poor
weather conditions (e.g. clouds) or observational difficulties
(e.g. variable vignetting of the seeing disk by the slit). Further
information is given in \v{S}imi\'c et al. (1998). In a last step the
trailed spectra were phase-folded into 125 phase bins using the updated
eclipse ephemeris as given in Sect. 3.1.

\begin{table}
\caption{\label{bright} Magnitudes $m_{max}$, $m_{min}$ and orbital
modulation $\Delta m$ of RXJ0019.8 on Oct 12,1996.} 
\begin{tabular}{cccc}
\hline\hline
Band & $m_{max}$ & $m_{min}$ & $\Delta m$ \\
\hline
U & 11.37 & 11.79 & 0.42 \\

B & 12.29 & 12.71 & 0.42 \\

V & 12.33 & 12.65 & 0.42 \\
\hline
\end{tabular}
\end{table}

%__________________________________________________________________

\section{Analysis and results}

\subsection{The lightcurves}

All observed MCCP and MEKASPEK lightcurves are presented in
Fig. \ref{lc}. Only B-band
lightcurves are plotted since the very blue continuum of RXJ0019 in
combination with the quantum efficiency of the detectors and the
color temperature of the comparison provides the best S/N value in
B. All observations are plotted individually to emphasize the high
variability of this object. A plot including all observed lightcurves
superposed according to the orbital period can be found in
Meyer-Hofmeister et al. (1998) in their Fig. 1. Also compare 
with Fig. 2 in Will \& Barwig (1996).
The following features can be observed:

%
%%%%%%%%%%%%%%%%%%%%%%%%%  Abbildungen Minima / Colorvariations %%%%%%%%
%
%
\begin{figure}
\includegraphics[width=\hsize]{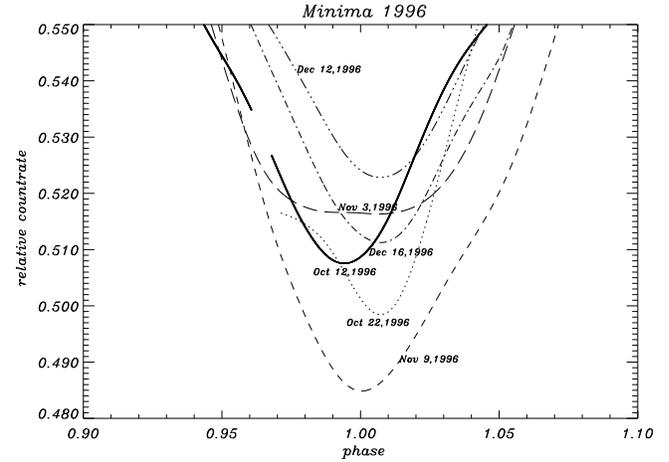}
\caption{Minima observed in 1996. The minima occur statistically around
phase $\phi=0.0$. The depth and the shape of the minima varies. All
lightcurves were approximated by splines.}
\label{minkur}
\end{figure}
\begin{figure}
\includegraphics[width=\hsize]{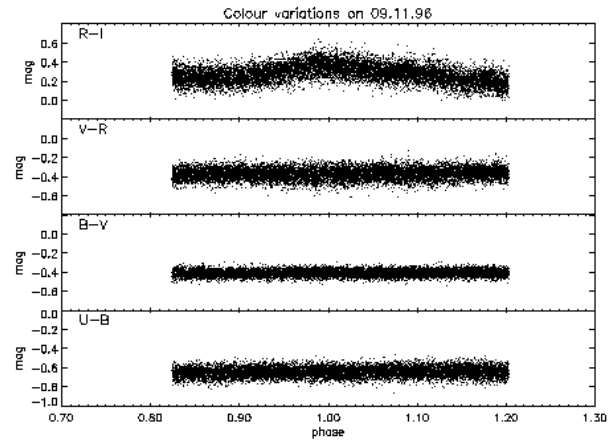}
%,bb=66 372 530 706]{colvar.ps}
\caption{Color Variations on Nov 9, 1996. Symmetrically to phase 0.0
an increase in the $R-I$ flux can be observed with an amplitude of
$\approx$\,$0.15^m$. The zero point of the vertical scale is arbitrary. }
\label{col}
\end{figure}
%
%
%
%%%%%%%%%%%%%          ABBILDUNG DER LICHTKURVEN            %%%%%%%%%%%%%
%
%
%
\begin{figure*}
\includegraphics[keepaspectratio=false, height=14.7cm,width=\hsize]{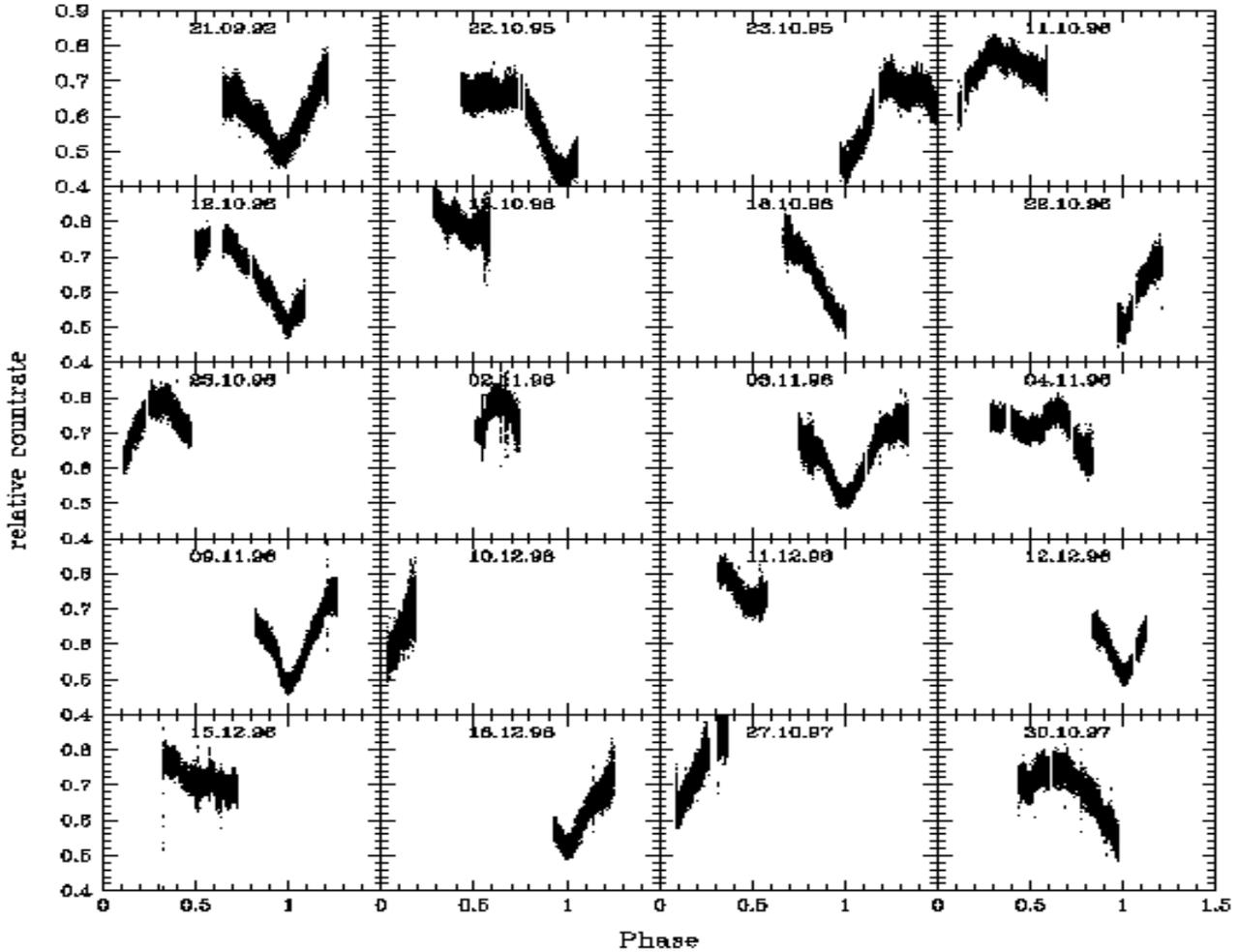}
% bb=38 178 567 695
\caption{MCCP and MEKASPEK B-band lightcurves between September 1992
to October 1997. Three lightcurves from 1992 and 1995 have also been
included (Will \& Barwig, 1996). Note that in 1995 the brightness of RXJ0019
is slightly lower. All lightcurves differ significantly from each
other. Also note the quasi-periodic 1.8h humps on Sept 21,1992, Oct
11,1996, Oct 13,1996. Increased scatter in the lightcurves is due to
very poor observing conditions. }
\label{lc}
\end{figure*}
%
%
% 
%%%%%%%%%%%%%%%%%%     Tabelle Minima    %%%%%%%%%%%%%%%%%%%%%%%%%%%%%%%% 
%
\begin{table}
\caption{\label{minima}Timings of primary minima (HJD)}
\begin{tabular}{cccc}
\hline\hline
date & cycle (E) & minima & mean \\
 & & HJD-2450000 [d] & error [d] \\
\hline
12/10/1996 & 22061 & 369.590 & 0.00099 \\ 
22/10/1996 & 22076 & 379.504 & 0.00175 \\
03/11/1996 & 22094 & 391.385 & 0.00774 \\
09/11/1996 & 22103 & 397.333 & 0.00065 \\
12/12/1996 & 22153 & 430.361 & 0.00130 \\
16/12/1996 & 22159 & 434.323 & 0.00036 \\
\hline
\end{tabular}
\end{table}

1.) RXJ0019 shows a deep, broad minimum ($\Delta m$\,=\,$0.42^m$ in
\em UBV \rm on Oct. 12, 1996, Table \ref{bright}), lasting about
0.55 of a complete orbital phase. As can be seen in
Fig. \ref{minkur} the depth and shape of the minima varies and the
timings of the minima are not regular. This explains the
systematic difference of the orbital period of Will \& Barwig (1996),
which lies out of the error bars of other periods given in the
literature (e.g. Greiner \& Wenzel 1995 or Matsumoto 1996): our
dataset was too short to eliminate the statistical effects. With the
epoch of the photometric minimum given by Greiner \& Wenzel (1995) and
the very accurate determination of the minima observed at the
Wendelstein observatory since 1992 we derive the following orbital
period:
\begin{center}
HJD = $2435799.247 + 0.6604573(40)\times$ E
\end{center}
The timings of the minima used for this calculation can be found in
Table \ref{minima} and in Will \& Barwig (1996) in their Table 2.

2.) A secondary minimum appears around phase $\phi\approx0.5$. The
amplitude of the secondary minimum is $\Delta m\approx 0.1^m$. The
following maximum at $\phi\approx 0.65$ is slightly fainter than the
first maximum. Sometimes the second maximum is even completely smeared
out and after a flat portion of the lightcurve the ingress to the main
minimum begins (e.g. Oct 22, 1995 and Dec 15, 1996 in Fig. \ref{lc}).

3.) We observe humps and step-like features during ingress and egress
of the main minimum. A period analysis of these features, when
apparent in the lightcurves, reveals a period of approximately 1.8\,h
(e.g. Oct 11,1996 and Oct 13,1996). In contrast to the lightcurves
showing steps and humps there are also lightcurves with a rather flat
appearance (e.g. Nov 09,1996 or Dec 12,1996). We assume that
RXJ0019 changes between two states: an ``excited optical state'' where
humps and steps appear quasi-periodically with a period of 1.8\,h and
a ``quiet optical state'', where the lightcurves show no additional
features.  RXJ0019 can change from one state into the other from night
to night.  A probable explanation might be a short-term variation in
the mass accretion rate (Meyer-Hofmeister et al. 1998).

%%%%%%%%%%%%%%%%%%%%%%  Abbildung Spektren %%%%%%%%%%%%%%%%%%%%%%%%%%%%%%%
%                   Mean orbital spectra.....
%
\begin{figure*}
\mbox{
\includegraphics[keepaspectratio=false,height=5cm,width=\columnwidth,bb
= 20 340 535 600]{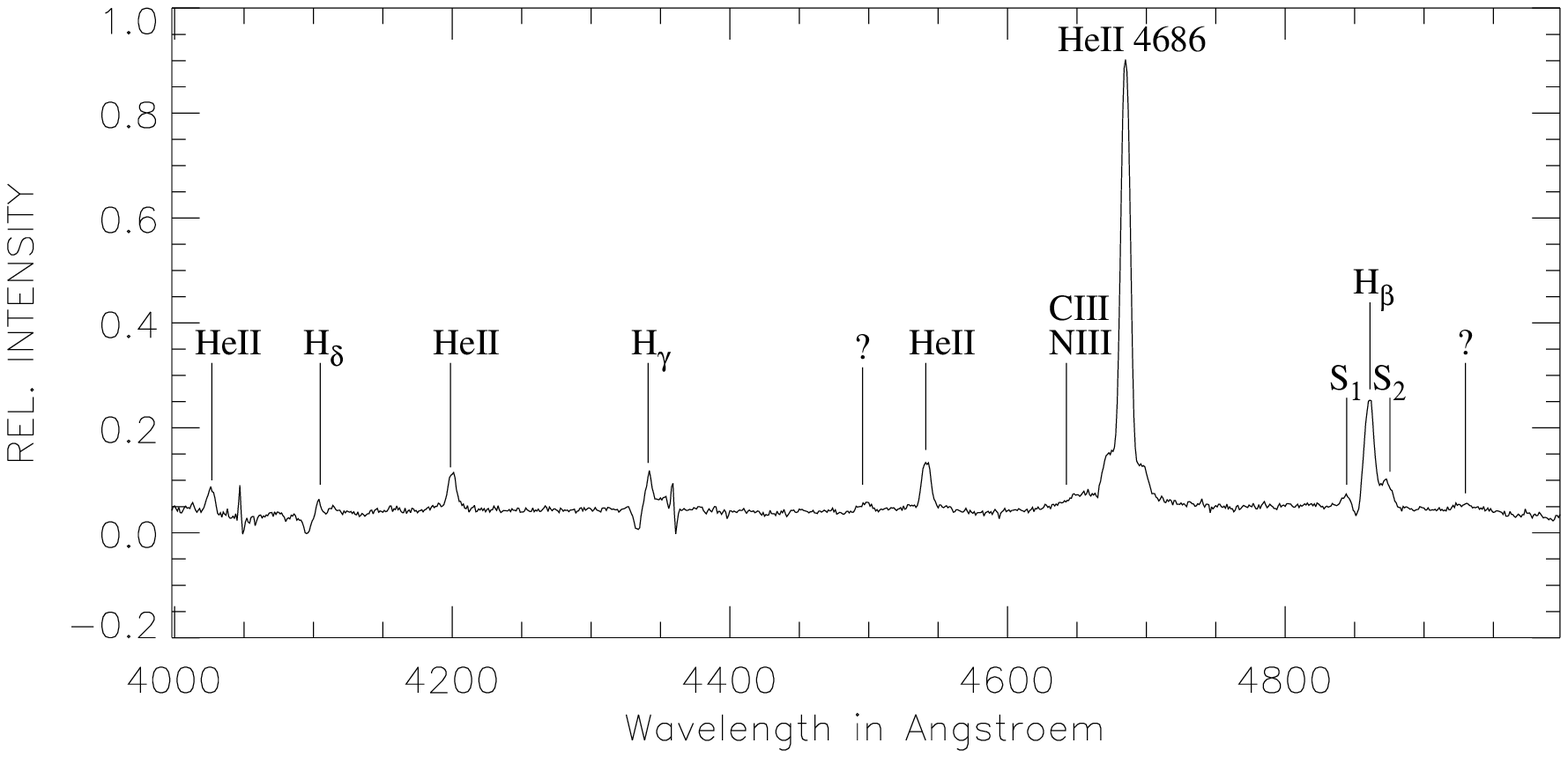}
\includegraphics[keepaspectratio=false,height=5cm,width=\columnwidth,bb
= 20 340 535 600]{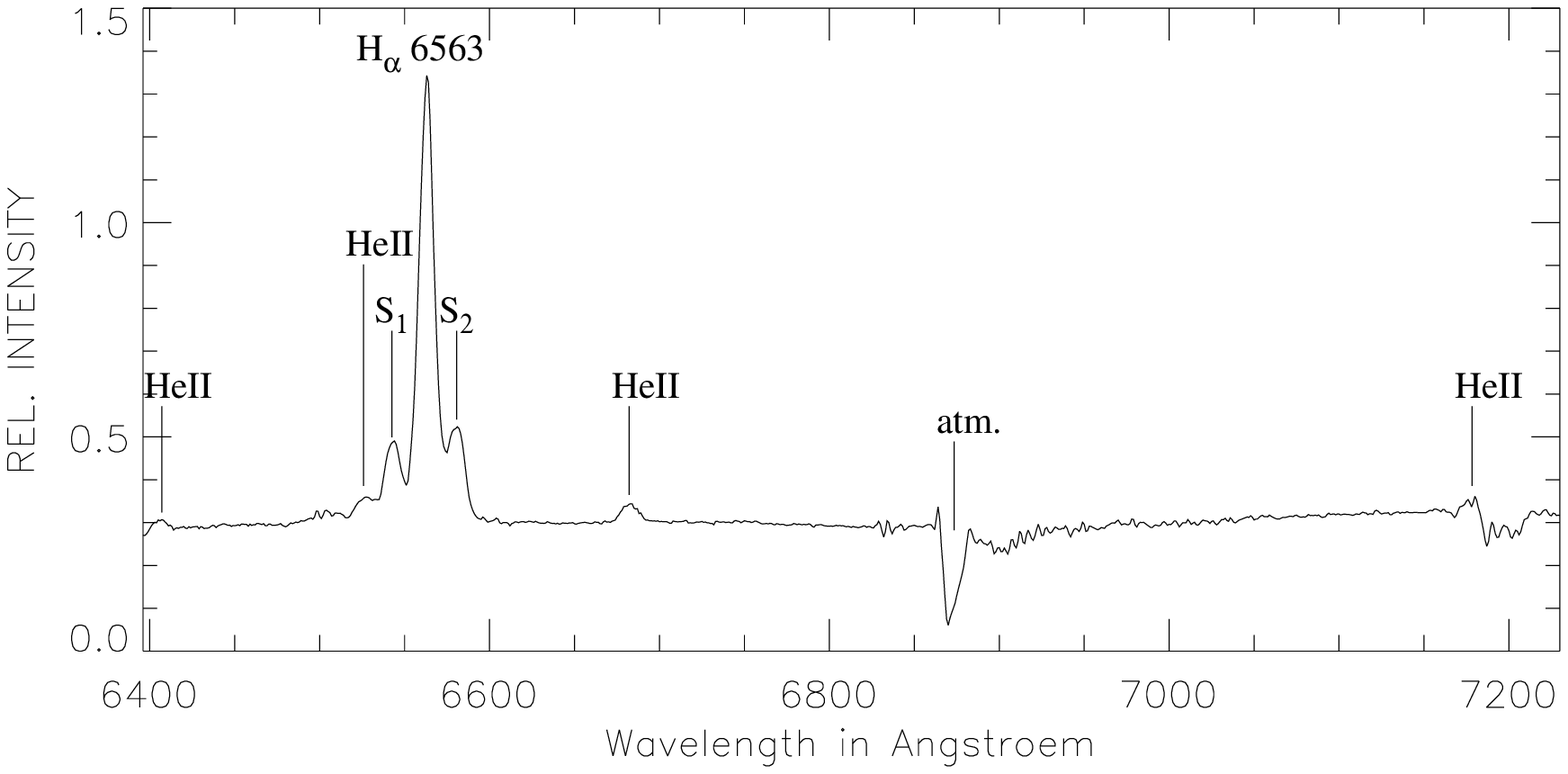}}
\caption{Mean orbital spectrum in the blue spectral range range from
4000{\AA} to 5000{\AA} and in the red spectral range from 6400{\AA} to
7250{\AA}. All Balmer lines show P-Cygni absorption on their blue wings.
The base of \ion{He}{ii} $\lambda$\,4686 is very broad and
asymmetric but the jet lines are not separated from the main emission.
The jet lines (labeled S$_1$ and S$_2$) at H$_{\alpha}$ and H$_{\beta}$
are  clearly visible.  H$_{\alpha}$ is considerably blended by two
\ion{He}{ii} emission lines.}
\label{meanspec}
\end{figure*}
%
%%%%%%%%%%%%%%%%%%%%%%%%%%%%%%%%%%%%%%%%%%%%%%%%%%%%%%%%%%%%%%%%%%%%%%%%%

4.) Matsumoto (1996) did not observe any color variations in his
lightcurves.  We also cannot find any color variations in $U-B$,
$B-V$ and $V-R$. But in $R-I$ we detect a weak flux increase
symmetrically to $\phi=0.0$ with an amplitude of $\Delta$m
$\approx0.15^m$. This variation in $R-I$ can only be observed when the
main minimum appears at midnight hours. In the evening and in the
morning the natural reddening of the sky prevents an observation of
this color variation. Therefore, we observed this color variation only
on Oct 12, Nov 3 and Nov 9, 1996 (Fig. \ref{col}). In every night
the width and the amplitude of this variation are similar. All other
portions of the color curves from phase $\phi=0.1$ to 0.9 are flat and
show no variations.

\subsection{The mean orbital spectra}

High-resolution spectra of RXJ0019 have recently been obtained by Tomov et
al. (1998) and Becker et al. (1998). We also present high resolution
mean orbital spectra (in Fig. \ref{meanspec}) in
the blue (3700...5000\AA) and in the red spectral range
(6300...8300\AA). Similar to other SSS (e.g. Crampton et al. 1996 and
Southwell et al. 1996) the very blue spectrum is dominated by
\ion{He}{ii} and Balmer lines. The Balmer series can be
traced up to H$_{12}$. All transitions of the
\ion{He}{ii} (n,3),(n,4) and (n,5) series can be observed when not
blended by Balmer lines. No \ion{He}{i} can be detected indicating a
high ionisation state.

Additionally, we see some higher ionization emission features such
as \ion{O}{vi} $\lambda$\,3811. The \ion{C}{iii}-\ion{N}{iii}
$\lambda\lambda$\,4640\,--\,4660 emission complex contributes to the blue
wing of the \ion{He}{ii} $\lambda$\,4686.  

%
%%%%%%%%%%%%%%%        Abbildung Spektren     %%%%%%%%%%%%%%%%%%%%%%%%%
%
%       2.Teil
%
\begin{figure}
\includegraphics[width=\columnwidth,bb=76 379 553 709]{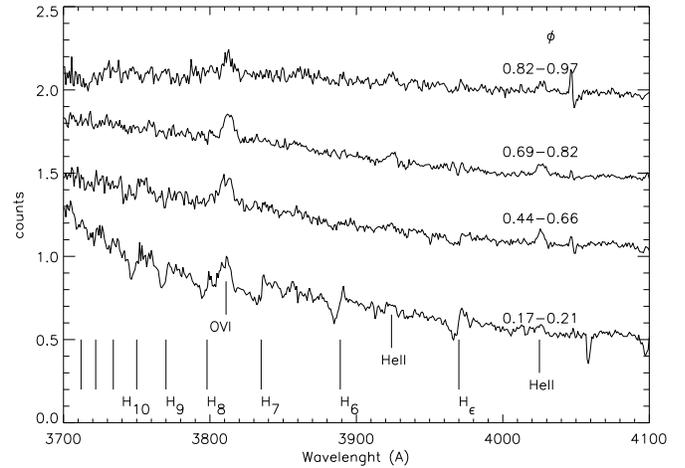}
\caption{Phase-resolved, uncalibrated  spectra showing the variability 
of the emission lines between 3700~{\AA} and 4100~\AA. The first
spectrum from $\phi=0.17-0.21$ is dominated by P-Cygni absorptions
at the Balmer lines up to H$_{12}$, which cannot be seen at later
phases. Marks at the bottom indicate the Balmer series up to H$_{13}$.   }
\label{bluebin}
\end{figure}
\begin{figure}
\includegraphics[width=\columnwidth,bb=76 379 553 709]{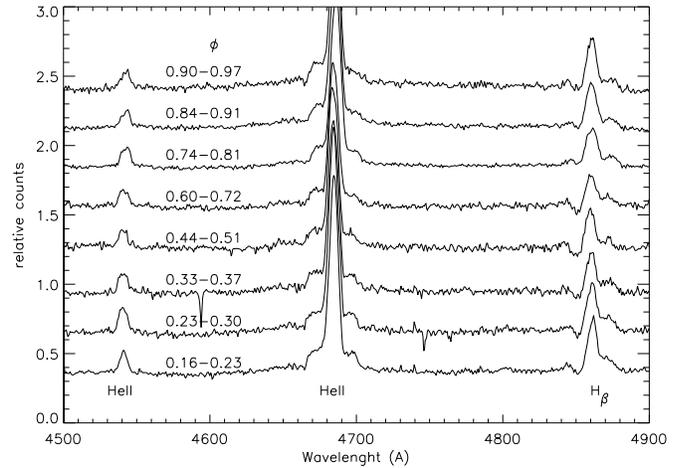}
\caption{Phase-resolved spectra showing the variability of the
emission lines between 4500 {\AA} and 4900 \AA. The P-Cygni absorption
at H$_{\beta}$ almost disappears between $\phi=0.7-0.9$. Also note the
change of the base of \ion{He}{ii} $\lambda$\,4686 on the red wing of
the emission line.}
\label{bin}
\end{figure}
%
%
%%%%%%%%%%%%%%%%%%%%%%%%%%%%%%%%%%%%%%%%%%%%%%%%%%%%%%%%%%%%%%%%%%%%%%%%

On the blue wings of the Balmer lines P-Cygni profiles are visible in
all observed lines. The H$_8$ to H$_{12}$ transitions can only be
detected by their corresponding P-Cygni absorption between $\phi=0.12$
to 0.30 (Fig. \ref{bluebin}). We do not observe these absorptions
after $\phi=0.30$.  At the other Balmer lines the P-Cygni absorption
shows an orbital modulation that almost disappears between phase
$\phi=0.7$ and 0.9 (Fig. \ref{bin}), indicating that there might be a
directed fast wind in this system. The velocity of the wind amounts to
$\approx$\,590\,kms$^{-1}$ with the blue absorption wing extending to
$\approx$\,900\,kms$^{-1}$. The \ion{He}{ii} emission lines are not
truncated by P-Cygni profiles.

Two emission features, which are clearly above the S/N ratio at
$\lambda\lambda$\,4500, 4930, could not be identified
with any reasonable ion species.

The Balmer lines H$_{\alpha}$, H$_{\beta}$ and H$_{\gamma}$ have
broad bases and show satellite lines symmetrically around the main
emission. All bright Balmer lines are blended by \ion{He}{ii} emission
lines, in particular the two Helium lines $\lambda\lambda$\,6527, 6560
significantly contribute to H$_{\alpha}$ and \ion{He}{ii}
$\lambda$\,4859 to H$_{\beta}$. Therefore, substructures in these lines are
blurred by the Helium emission, whereas in H$_{\gamma}$,
H$_{\delta}$ and H$_{\epsilon}$ similar complex substructures of
different components can be detected. For illustration the trailed
H$_{\gamma}$ spectrum is included in Fig. \ref{trail}.

In \ion{He}{ii} $\lambda$\,4686 we only observe a broad base. The
adjacent satellite components are not clearly separated from the main
emission and are asymmetric (Fig. \ref{meanspec} and \ref{bin}).
Tomov et al. (1998) and Becker at al. (1998) argue that these satellite
lines are the spectral signatures of jets originating near the white
dwarf. These collimated high velocity outflows have recently also been
reported by Quaintrell \& Fender (1998) from their infrared spectroscopy.

Transient jets seem to be a common feature among the SSS. Jets have
also been detected in RX\,J0513.9-6951 (e.g. Southwell et al. 1996)
and recently in RX\,J0925.7-4758 (Motch 1998). The observed projected
high outflow velocities of these sources, which are seen at very low
inclination, indicate an origin of the jet near the white
dwarf. Therefore, the jet velocity is close to the escape velocity of
the central object (see  Livio 1998) and can be used for estimating
the inclination of the system.

The projected velocities of the jet lines (measured at maximum of
their emission profiles) of RX\,J0019.8 are rather low. We measure 920
km\,s$^{-1}$ for S$_1$ and 805 km\,s$^{-1}$ for S$_2$. The asymmetry
can be explained by the P-Cygni absorption, which alters the appearance
of the jet emission on the blue side significantly. The low velocities
may be due to a high system inclination (Tomov et al. 1998) if the jet
originates near the white dwarf. A medium to high inclination is also
consistent with the deep eclipse lightcurves from our photometry. 

We note, that no spectral features of a secondary star are detected in
our high resolution spectra.

\subsection{Radial velocities}

%
%%%%%%%%%%%%%%%%%%%%%%     radial velocities      %%%%%%%%%%%%%%%%%%
%
\begin{figure}
\includegraphics[height=\columnwidth,angle=90,bb=56 320 358 738]{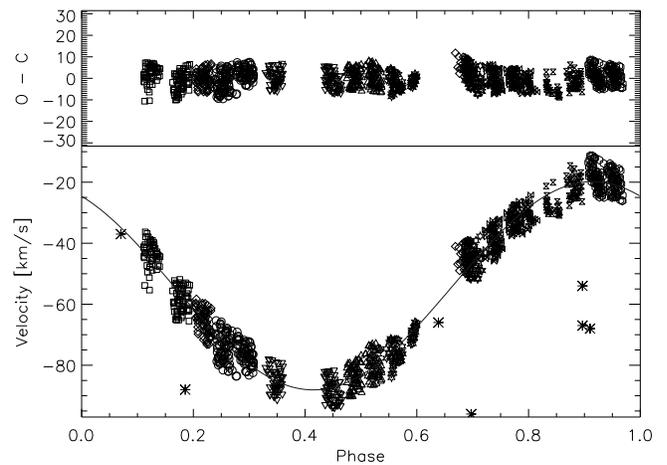}
\caption{Radial velocity curve determined by fitting Gaussians to the
trailed spectra of H$_{\alpha}$. The upper panel shows the residuals,
the lower the distribution of the data. The solid line is the sine-fit
to the data. For comparison the radial velocities from Tomov et
al. (1998) are also plotted as asterisks (different symbols of our
measurements refer to different trailed spectra).}
\label{harv}
\end{figure}
\begin{figure}
\includegraphics[height=\columnwidth,angle=90,bb=56 320 358 738]{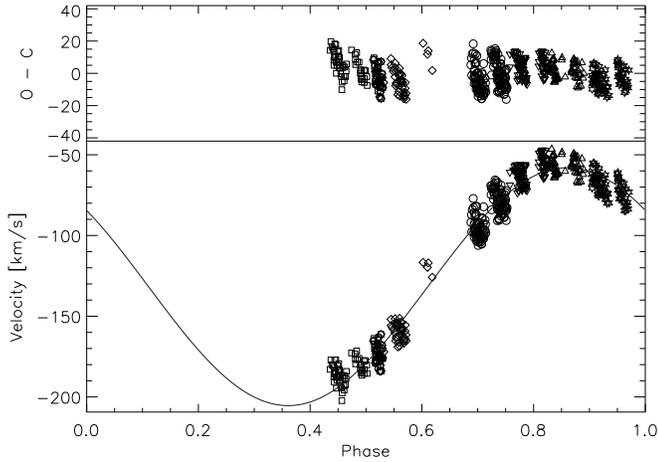}
\caption{Radial velocity curve determined by fitting Gaussians to the
trailed spectra of \ion{He}{ii}. The upper panel shows the residuals,
the lower the distribution of the data. The solid line is the sine-fit
to the data. As the CCD is slightly warped we measure a systematic
shift of the radial velocities within a single spectrum (see text) to
higher velocities.}
\label{herv}
\end{figure}
%
%
%%%%%%%%%%%%%%%%%%%%%%%%%%%%%%%%%%%%%%%%%%%%%%%%%%%%%%%%%%%%%%%%%%%%%%%

The usual procedure of measuring radial velocities in complex emission
lines is fitting Gaussians to their line profiles.  Measuring radial
velocities of the emission lines of the RXJ0019 spectra is
difficult. First, the Balmer emission lines are truncated by the phase
dependent P-Cygni profiles and blended by \ion{He}{ii} emission lines
which will alter the system velocity and semi amplitude
considerably. Second, the asymmetric profile of the satellite lines
will cause systematic errors in the velocity
determinations. Furthermore, the measured velocities are only of the order
of the spectral resolution of the spectrograph.  Therefore, we
restricted our investigation to the strongest emission lines
H$_{\alpha}$ and \ion{He}{ii}.

For the pixel rows of the central parts of our trailed spectra we fitted
Gaussians to the emission line profile. In order to suppress the
influence of the noise and the satellite lines, the central region of the
Gaussians was given a bigger weight than the adjacent spectral range
(Bevington \& Robinson 1992). After that a sine-fit was applied to the
fitted radial velocity $V$ and the observed semi-amplitude $K_{obs}$,
using the following relation:

\begin{equation}
V = \gamma + K_{obs} \sin(\phi - \phi_{0})
\end{equation}
where $\gamma$ is the system velocity, $K_{obs}$ is the radial
semi-amplitude of the white dwarf (assuming these emission lines
originate near the primary component), and $\phi$ and $\phi_{0}$ indicate the
phase and the phase offset of the binary system, respectively. 

As we do not have blue calibration spectra from the first observing night our
velocity determinations of \ion{He}{ii} were restricted to the spectra
obtained in the last night. 

There is a systematic shift in our velocity measurements: measurements
from a trailed spectrum show a continuous drift to higher
velocities. The reason for this is that the CCD chips are slightly
warped. We account for this effect by fitting sky lines in the spectra
from which we calculate the distortion. But the effect could not be
eliminated in the blue spectral range, as there were only two weak sky
lines which could not be used for a reasonable fit. Therefore, no
correction was made in the blue spectral range. It should be noted
that in spite of this systematic error our residuals are still 
low (Figs. \ref{harv} and \ref{herv}). 

%%%%%%%%%%%%        mass function     %%%%%%%%%%%%%%%%%%%%%%%%%%%%%%%
%
\begin{figure}
\includegraphics[width=\columnwidth]{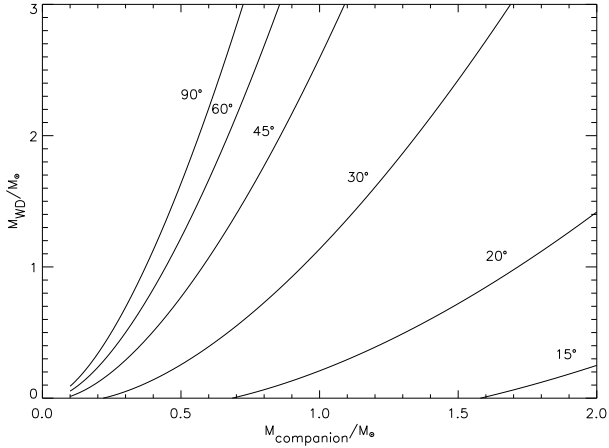}
\caption{Primary mass of RXJ0019 for various inclinations, $i$,
plotted from the mass function $f(M)=0.0274 M_{\sun}$. For any
reasonable inclination it is clear that the companion must be a
low-mass star.}
\label{massf}
\end{figure}
%
%%%%%%%%%%%%%%%%%%%%%%%%%%%%%%%%%%%%%%%%%%%%%%%%%%%%%%%%%%%%%%%%%%%%%

The sine-fits to the radial velocity curves of H$_{\alpha}$ and
\ion{He}{ii} are given in Figs. \ref{harv} and \ref{herv},
respectively. The deduced velocities are listed in Table \ref{velo}.
It is obvious, that there is a substantial difference between the
radial velocities in H$_{\alpha}$ and \ion{He}{ii}: the semi-amplitude
in \ion{He}{ii} is considerably higher. The difference in the radial
velocity between the two lines  can be explained by
the blurred H$_{\alpha}$ emission (see Sect. 3.2).  Our result for the radial
velocity of \ion{He}{ii} $\lambda$\,4686 is $73.7\pm 2.6$
km\,s$^{-1}$, which is in accordance with other values given in the
literature (Becker et al. 1998, Beuermann et al. 1995). 

We also observe a considerable difference in the system velocity
between H$_{\alpha}$ and \ion{He}{ii} in our data. This effect is
caused by the P-Cygni profiles: the blue wing of the H$_{\alpha}$
emission is truncated by the P-Cygni absorption and shifts the maximum
of emission to slightly longer wavelengths. Thus, the system velocity
in H$_{\alpha}$ differs notably from the velocity measured in
\ion{He}{ii}. Compared with other values published in literature
(Becker et al. 1998) our derived system velocities are much higher,
especially for \ion{He}{ii}.  For a reliable determination of the
system velocity a dataset covering only a single orbital period is
probably too short to eliminate statistical effects, which are caused
by the varying influence of the jet emission.  For comparison we show
in Fig. \ref{harv} the velocities given in Tomov et al. (1998) in
their Table 1. It is obvious that their system velocity for
H$_{\alpha}$ is even higher.  But due to the systematic shift in our
distorted spectra our measured system velocity should not be given too
much weight.

Photometric minimum and spectroscopic phase zero do not coincide.  The
photometric minimum occurs 0.11 and 0.17 in phase before spectroscopic
phase zero for \ion{He}{ii} and H$_{\alpha}$, respectively.  A similar
shift for \ion{He}{ii} was already observed by Beuermann et
al. (1995).

For further investigations we adopt our results of the semi-amplitude
for the rest of this paper. The system velocities for the Doppler
tomography are taken from Becker et al. (1998).

As the jet lines of H$_{\alpha}$ mimic the orbital velocity modulation
of \ion{He}{ii} (Fig. \ref{trail}), we
assumed that the radial velocity modulation of \ion{He}{ii} is related
to the primary and therefore represents the motion of the white dwarf.
Knowing the orbital period and amplitude of the radial velocity
variations of the mass accretor we can deduce the mass function of the
secondary star. For the calculation we use the data from the
\ion{He}{ii} emission line. The mass function is determined by

%
%%%%%%%%%%%%%%%%%%%%%%%%% Gleichung Massfunction %%%%%%%%%%%%%%%%%%%%%%%%%%
%
\begin{equation}
f_{opt}(M_2) = \frac{P_{orb}K_1^3}{2\pi G} = \frac{{M_2^3}\sin^3
i}{(M_1 + M_2)^2} = 
0.0274(28) M_{\sun}
\end{equation}
%
%%%%%%%%%%%%%%%%%%%%%%%%% Tabelle Radialgeschwindigkeiten %%%%%%%%%%%%%%%%%
%
%
\begin{table}
\caption{\label{velo} Orbital parameters for RXJ0019. $K_{obs}$ is the
amplitude of the radial velocity and $\gamma$ denotes the system
velocity in km\,s$^{-1}$. $\phi_0$ denotes the phase offset between
photometric and spectroscopic phase (see Text). $\sigma$ designates
the errors of the mentioned parameters. The system velocities are not
consistent within their error bars. See text for an explanation.}
\begin{tabular}{ccccccc}
\hline\hline
Line & $K_{obs}$ & $\sigma_{K_{obs}}$ & $\gamma$ & $\sigma_{\gamma}$ &
$\phi_0$ & $\sigma_{\phi_0}$ \\
\hline
\ion{He}{ii} & 73.7 & 2.6 & -131.7 & 25.0  & 0.11 & 0.03   \\
H$_{\alpha}$ & 33.0 & 1.8 & -53.0  & 12.0  & 0.17 & 0.01   \\
\hline
\end{tabular}
\end{table}
where $P_{orb}$ is the orbital period, $K_1$ is the radial velocity of
the primary, $M_1$ and $M_2$ denote the mass of the white dwarf and
the secondary, respectively, and $i$ is the inclination.

The mass function is obviously very low. If the emission lines of
\ion{He}{ii} are indeed related to the compact object we can deduce
the mass of the companion. In Fig. \ref{massf} we plot our derived
mass function for various inclinations $i$.  Because of the arguments
given in Sect. 3.2, we assume that the inclination is reasonably
high. We conservatively estimate that the inclination ranges between
$50^{\circ}<i<90^{\circ}$.  If the compact object is a white dwarf
with $0.6<M_{1}<1.2 M_{\sun}$, then we obtain $0.3<M_{2}<0.5
M_{\sun}$ for the mass of the companion. The companion star is almost
certainly a low-mass object. This could explain why we do not see any
spectral features of the companion in our high resolution spectra. As
low mass functions are also known from other SSS (e.g. Crampton et
al. 1996, Hutchings et al. 1998), a low mass companion star may be
common among this class of objects. A possible evolutionary scenario
for low-mass secondaries in SSS is proposed by van\,Teeseling \& King (1998).

\subsection{Doppler tomography}

%
%
%%%%%%%%%%%%%           Trailed spectra        %%%%%%%%%%%%%%%%%%%
%
%
\begin{figure*}
\mbox{
  \includegraphics[angle=270,width=\columnwidth]{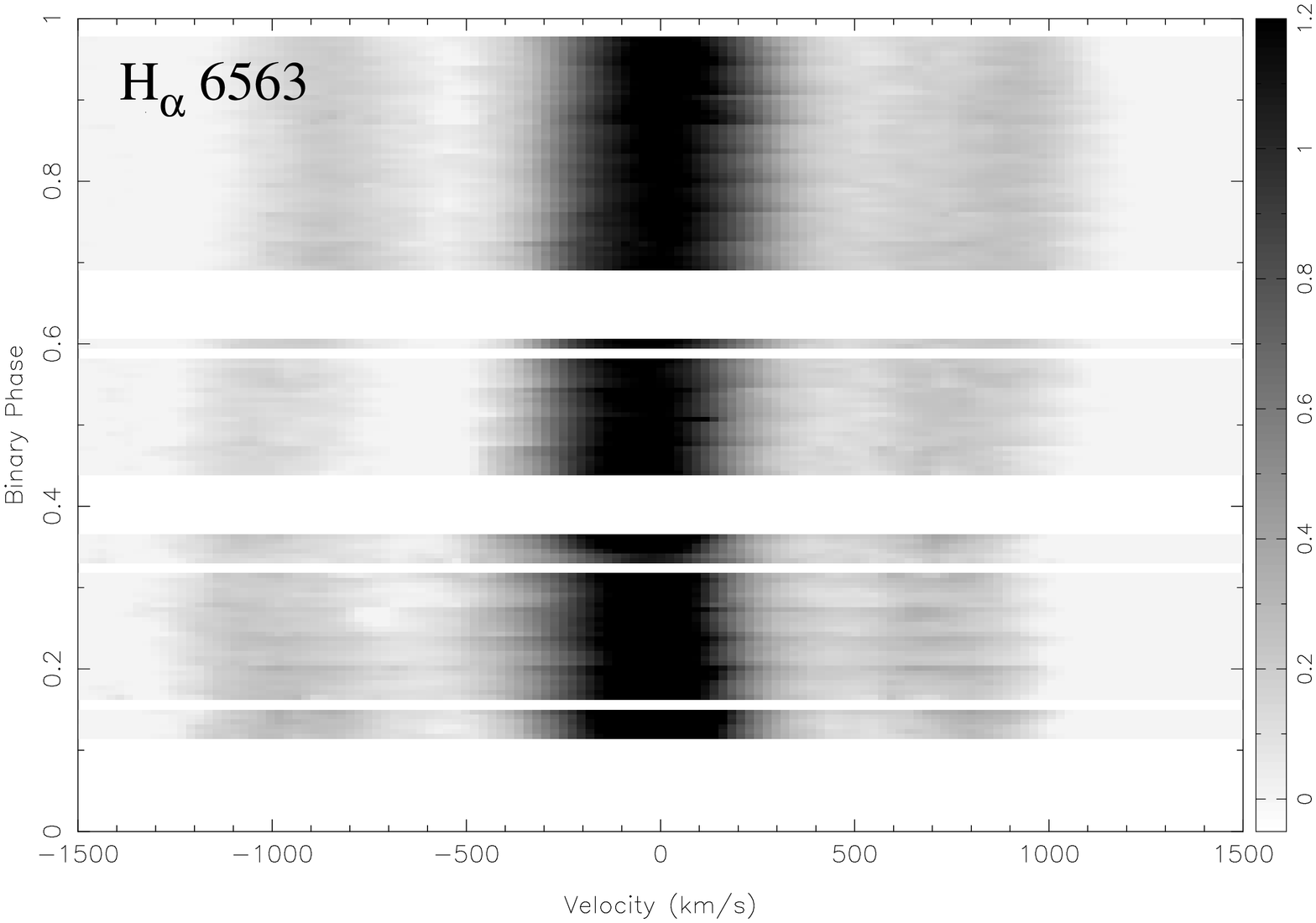}
  \includegraphics[angle=270,width=\columnwidth]{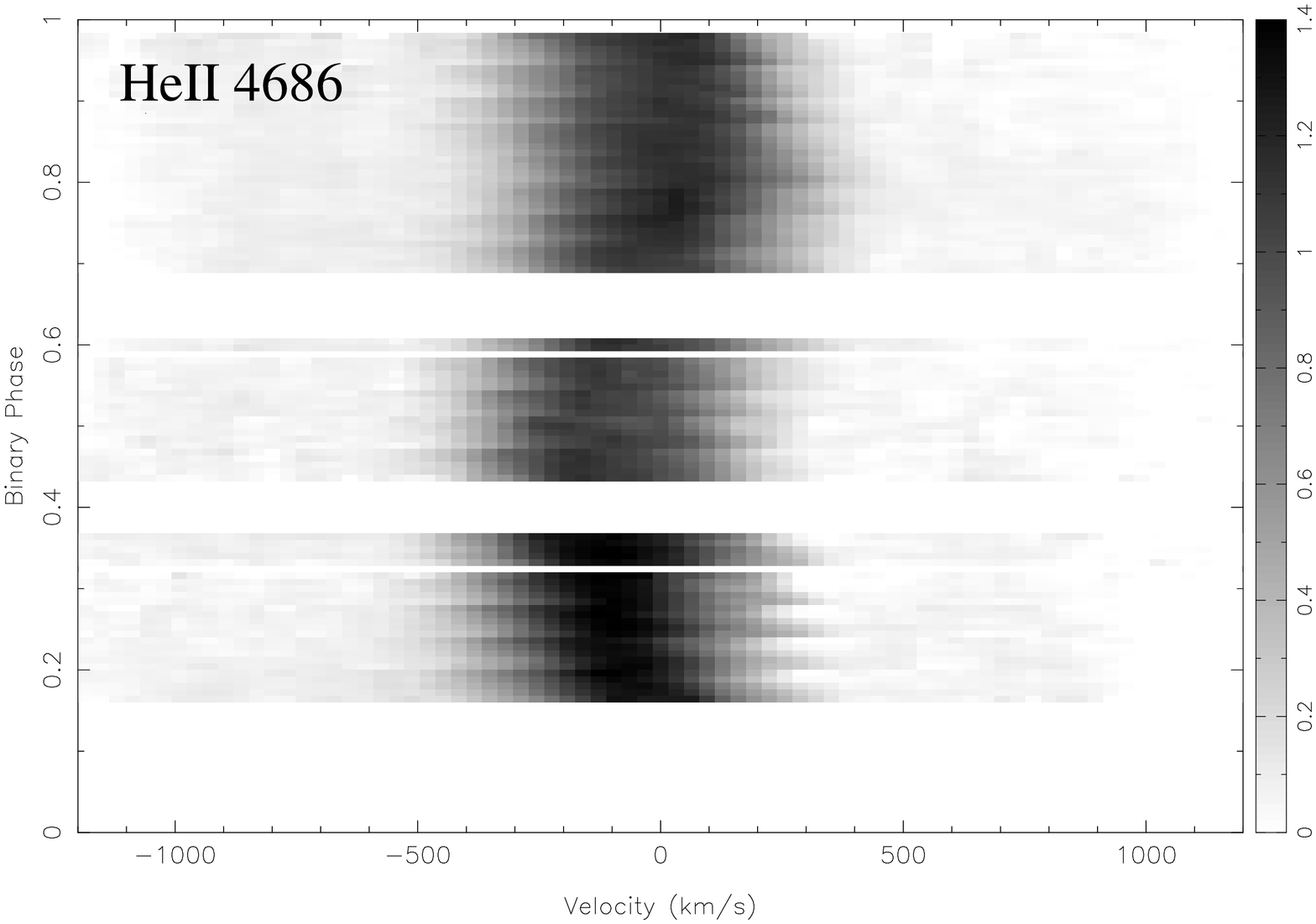}}
\mbox{
  \includegraphics[angle=270,width=\columnwidth]{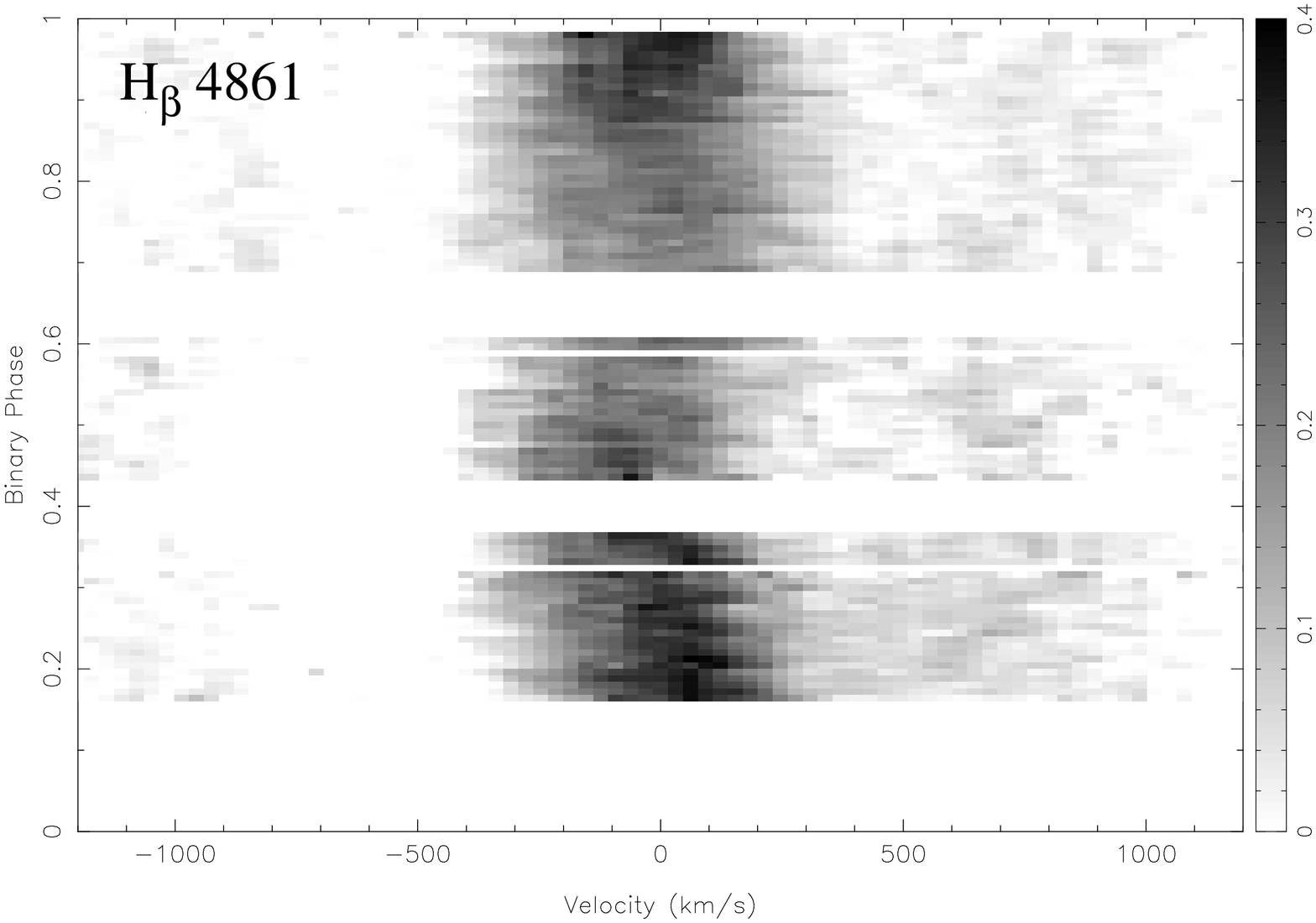}
  \includegraphics[angle=270,width=\columnwidth,]{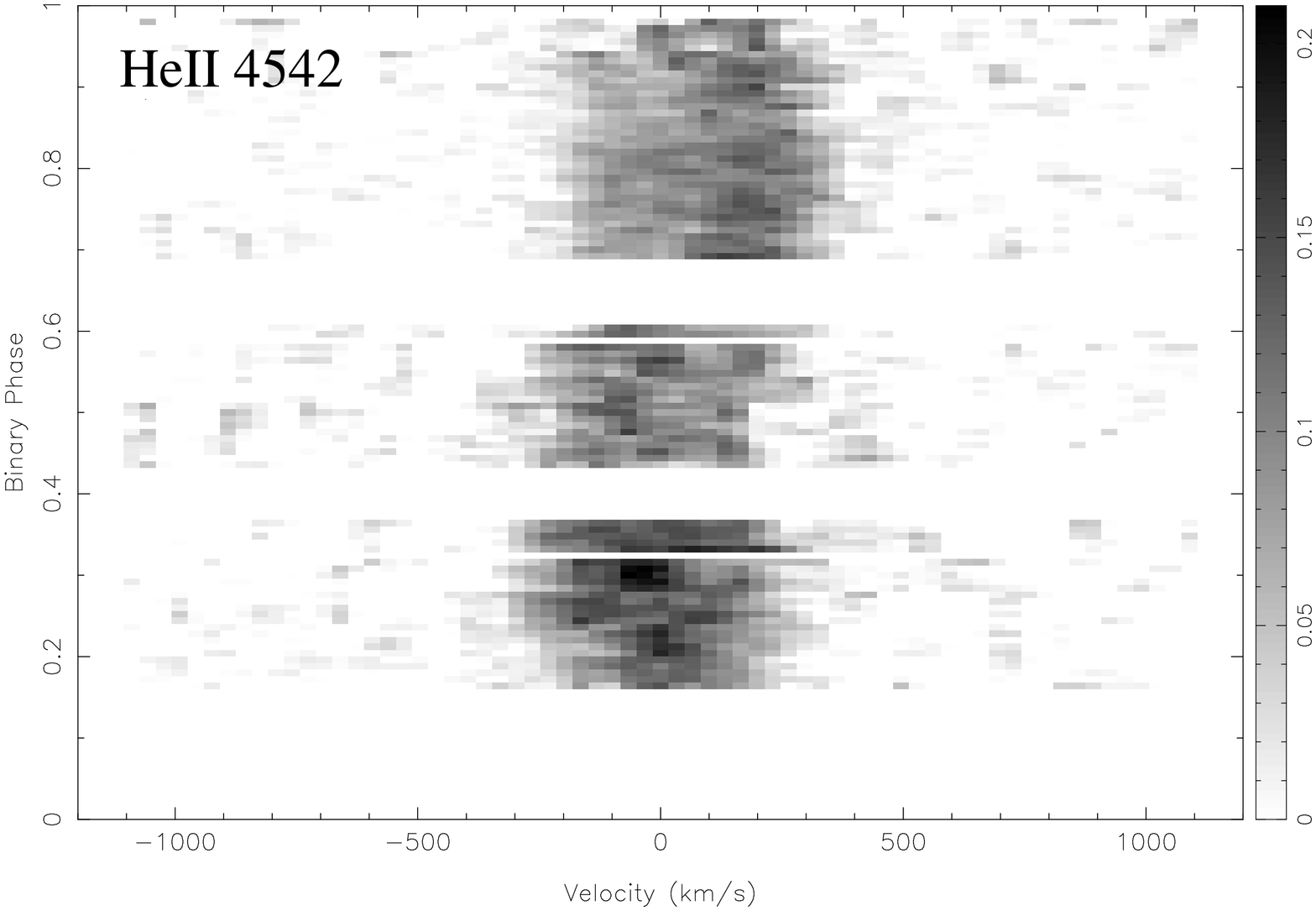}}
\resizebox{\columnwidth}{!}{
  \includegraphics[angle=270,width=\columnwidth,]{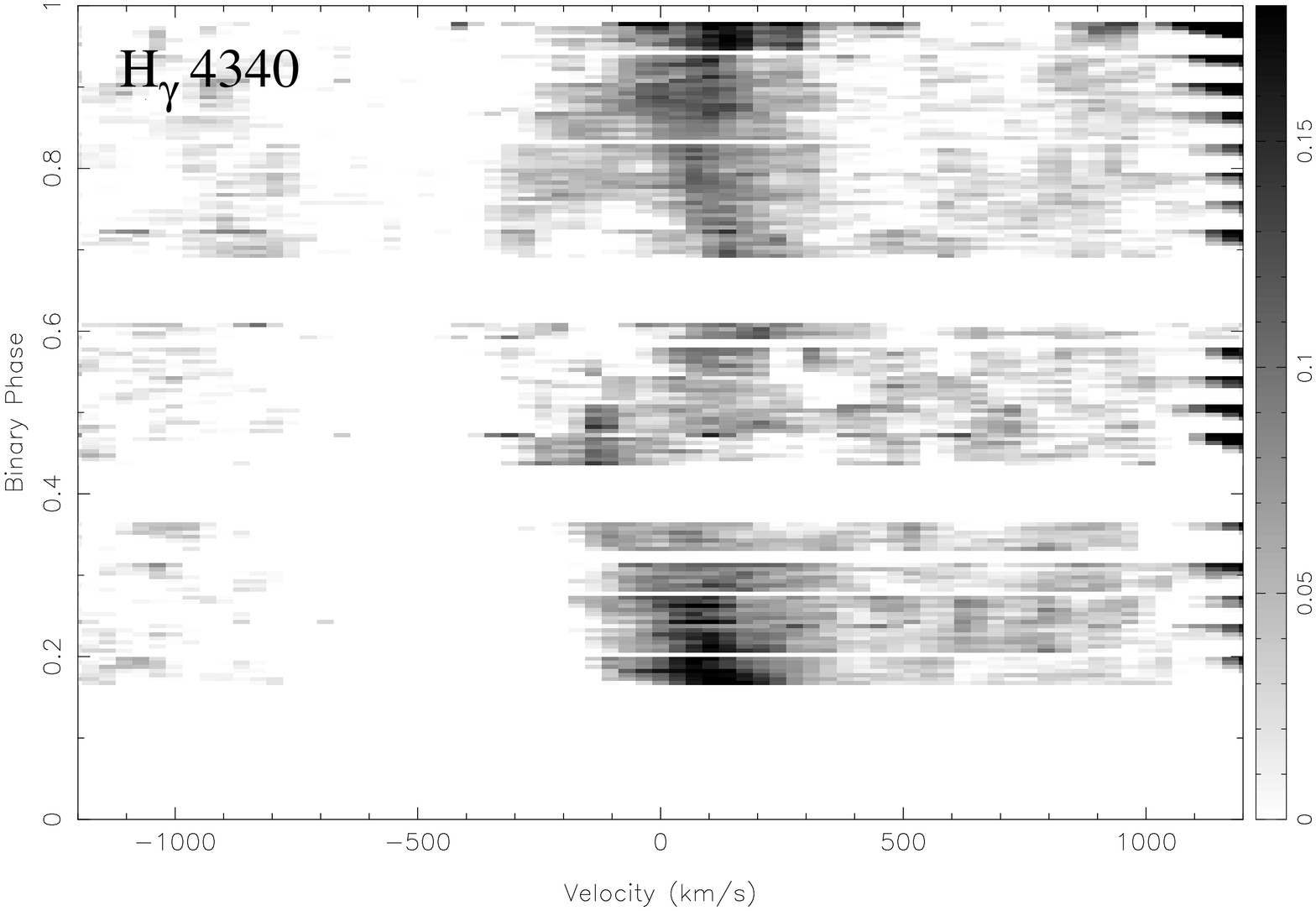}}
\hfill
\parbox[t]{\columnwidth}{\vspace{1.5cm}
  \caption[]{\label{trail}
  Zoomed grey scaled representations of continuum subtracted,
  photometrically calibrated, high resolution spectra of H$_{\alpha}$,
  H$_{\beta}$, H$_{\gamma}$ (left side), \ion{He}{ii} ${\lambda}$\,4686
  and \ion{He}{ii} ${\lambda}$\,4542 (right side) in v,
  $\phi$-coordinates. Not covered phases are
  marked with empty rows. In H$_{\alpha}$ the jets are clearly visible
  and show an orbital motion similar to the \ion{He}{ii} emission
  lines, whereas the main emission shows almost no orbital motion.
  The blue side of the Balmer lines are truncated by the
  P-Cygni absorption. In H$_{\gamma}$ complex substructures can be
  detected. The fragmented line appearing in the H$_{\gamma}$ spectra
  at a velocity of 1200 kms$^{-1}$ is a remnant of a night sky line.
  Within the Helium lines a weak S-wave component is visible.}}   
\end{figure*}
%
%
%%%%%%%%%%%%%%%%%%%%%%%%%%%%%%%%%%%%%%%%%%%%%%%%%%%%%%%%%%

Doppler tomography is an useful tool to extract further information on
the emission line origin from trailed spectra. This indirect imaging
technique, which was developed by Marsh \& Horne (1988), uses the
velocity of emission lines at each phase to create a two-dimensional
intensity image in velocity space coordinates $(V_X,V_Y)$.
The Doppler map can be interpreted as a projection of
emitting regions in accreting binary systems onto the plane
perpendicular to the observer's line of sight. 
The Doppler map is a function
of the velocity $(V_X,V_Y)$, where the $X$-axis points from the white
dwarf to the secondary and the $Y$-axis points in the direction of the
secondary's motion. 

An image pixel with given velocity coordinates
$(V_X,V_Y)$ produces an S-wave with the radial velocity
\begin{equation}
V = \gamma - V_X \cos(2\pi\phi) + V_Y \sin(2\pi\phi)
\end{equation}
where $\gamma$ and $\phi$ denote the system velocity and the
phase, respectively.
%
%
%
%%%%%%%%%%%%%%               TOMOGRAMME             %%%%%%%%%%%%%%%%%%%%%
%
%
\begin{figure*}
\begin{center}
\mbox{ 
  \includegraphics[width=\columnwidth,bb=41 114 572 630]{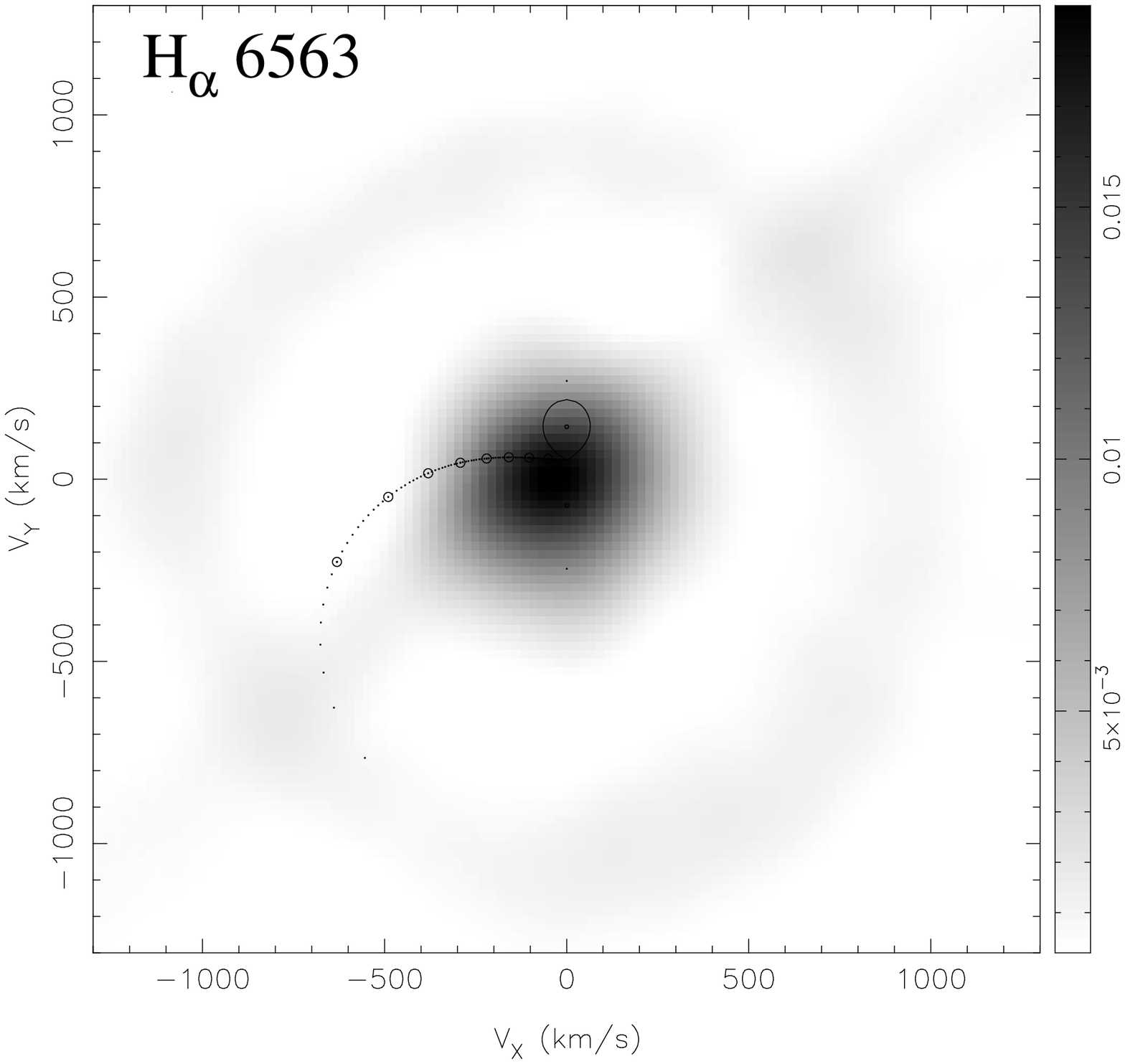} 
  \includegraphics[width=\columnwidth,bb=41 114 572 630]{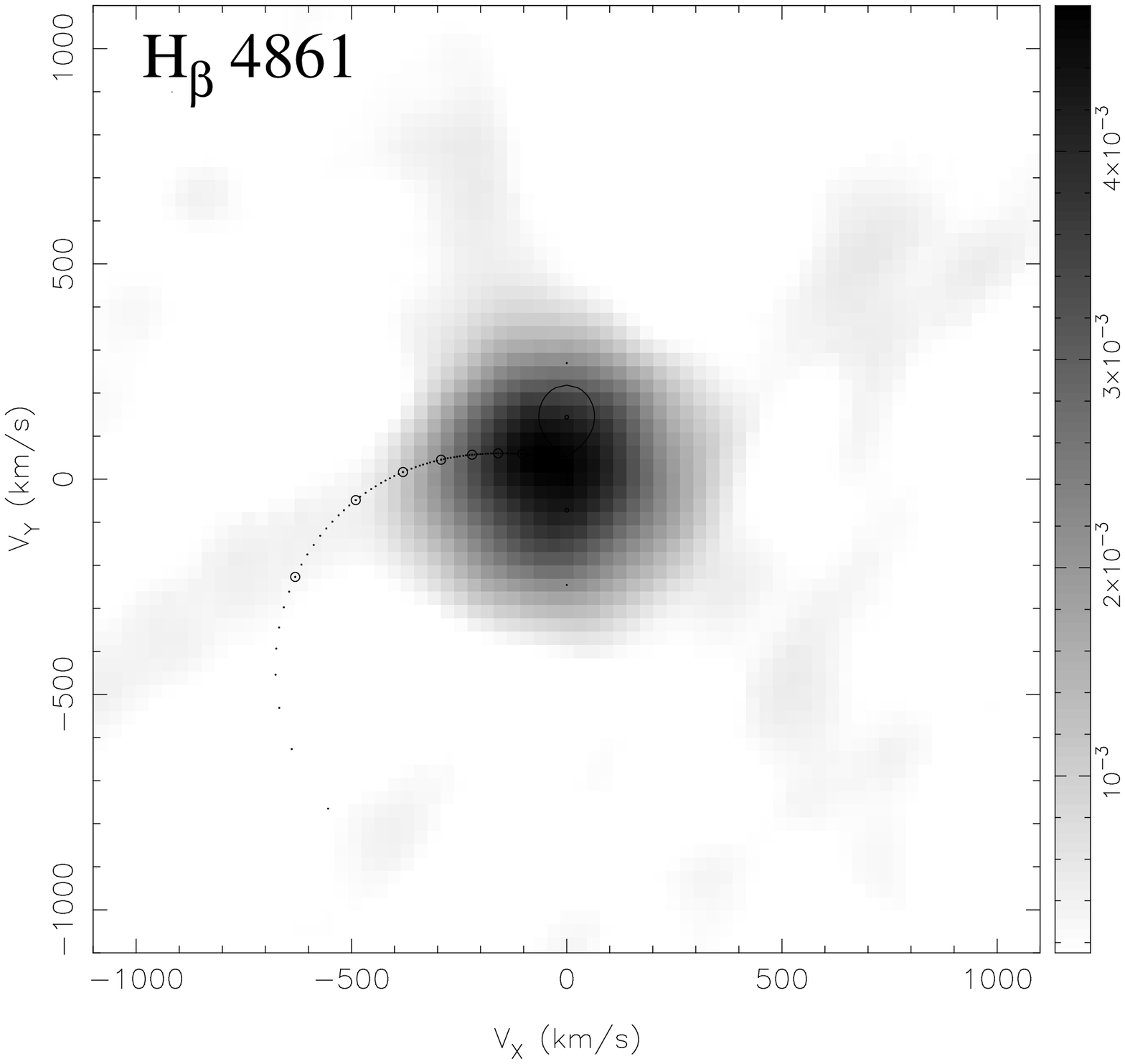}} 
\mbox{ 
  \includegraphics[width=\columnwidth,bb=41 114 572 630]{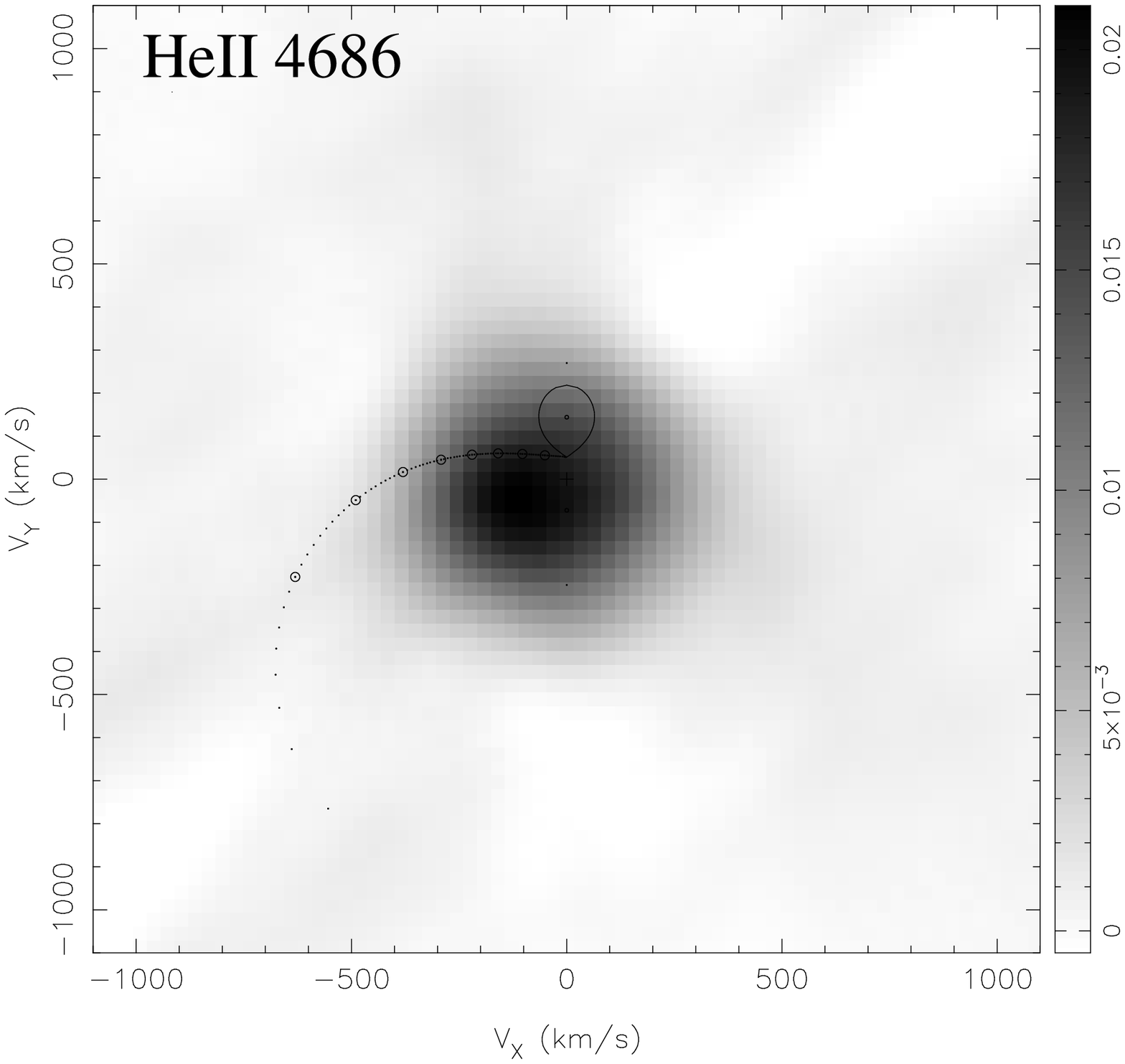} 
  \includegraphics[width=\columnwidth,bb=41 114 572 630]{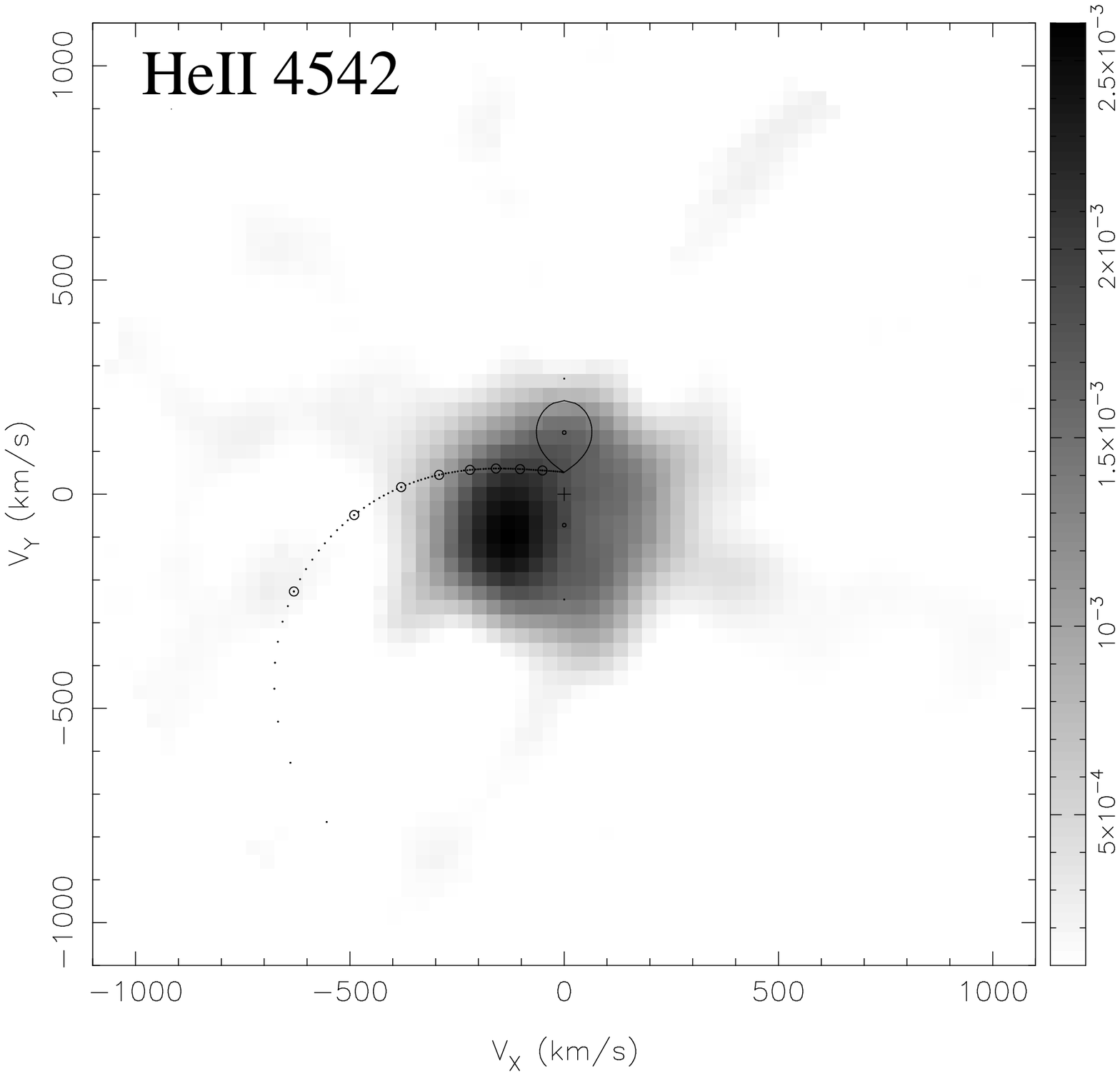}} 
\caption[]{Doppler--maps of H$_{\alpha}$, H$_{\beta}$, \ion{He}{ii}
  ${\lambda}$\,4686 and \ion{He}{ii} ${\lambda}$\,4542 in velocity
  space ($V_X,V_Y$), using the Fourier-filtered back-projection
  technique. The emission mainly originates in a ``disk'' centered at
  the center of mass, which extends to velocities of only 350
  kms$^{-1}$. The maximum of the emission of the Helium lines is shifted to
  $V_X \approx$ --100 kms$^{-1}$ and $V_Y \approx$ --100 kms$^{-1}$.
  The dim ring in H$_{\alpha}$ is caused by the jet emission. From the
  distribution of the emission in the Doppler maps it is clear, that
  the emitting material is not located in an accretion disk.  }
\label{tomo}
\end{center}
\end{figure*}
%
%%%%%%%%%%%%%%%%%%%%%%%%%%%%%%%%%%%%%%%%%%%%%%%%%%%%%%%%%%%%%
%

To accomplish this, a linear tomography algorithm, the
Fourier-filtered back-projection (FFBP) is used, which is described in
detail by Horne (1991). The resulting Doppler map (or tomogram)
is displayed as a grey-scale image. To assist in
interpreting Doppler maps we also mark the
position of the secondary star and the ballistic trajectory of the gas
stream. Assuming a Keplerian velocity field, emission originating
in the inner parts of the disk has a larger velocity and thus appears
in the outer regions of the map. Therefore, an image in such a
representation is turned inside-out.
For the tomogram analysis we restricted our investigation to the two
Balmer lines H$_{\alpha}$ and H$_{\beta}$ and the emission lines
\ion{He}{ii} ($\lambda$\,4686) and \ion{He}{ii}
($\lambda$\,4542) . No reasonable Doppler maps could be produced
for other emission lines because of their poor S/N ratios. 

Before computing the tomograms we subtracted the underlying continuum
from the individual emission lines, since their line flux is the
quantity needed to produce these maps. This was done by subtracting
from each pixel row of the trailed spectra the corresponding median of
the intensity.
We achieved a FWHM resolution of 150
kms$^{-1}$ in the central regions of the tomograms, whereas at higher
velocities ($\approx$\,800\,kms$^{-1}$) the resolution of the maps
suffers a considerable degradation.
Linear structures in the high velocity regions (Fig. \ref{tomo}) of
the maps are sampling artifacts (aliasing streaks) which are not taken
into account in the further interpretation. For a detailed discussion
about sampling artifacts see e.g. Robinson et al. (1993).

In Fig. \ref{trail} we present the trailed spectra of the emission
lines in V, $\phi$-coordinates; the Doppler tomograms
are shown in Fig. \ref{tomo}. The schematic overlays in the Doppler
maps represent the Roche-lobe of the companion star and the gas
stream. They are a function of $K_1$ and $K_2$. From our
discussion in Sect. 3.3 we have taken 1$M_{\sun}$ for the white dwarf
and a 0.5$M_{\sun}$ companion star. The center of mass and the
location of the white dwarf are respectively marked by a cross and a point
below the Roche lobe. The ballistic stream is represented by
an arc originating from the secondary's Roche lobe at the inner Lagrangian
point. This arc is marked every 0.1$R_{L_1}$ (open circles, $R_{L_1}$ is
the distance from the center of the primary to the inner Lagrangian
point) as it accelerates towards the compact object.

In the Balmer emission lines the blue side is truncated by the P-Cygni
absorption. No S-wave structures can be detected at H$_{\alpha}$ and
H$_{\beta}$.  The jet lines of H$_{\alpha}$ clearly show an orbital
motion which is in phase with the motion of \ion{He}{ii}. The main
emission component of H$_{\alpha}$ reveals almost no orbital motion
(compare Table \ref{velo}). Therefore, we assume that the jet and the
main emission do not have the same spatial origin.

The trailed spectra of both \ion{He}{ii} lines show more details:
in each spectrogram a weak S-wave component within the main emission
can be detected.

\begin{figure}
\includegraphics[width=\columnwidth]{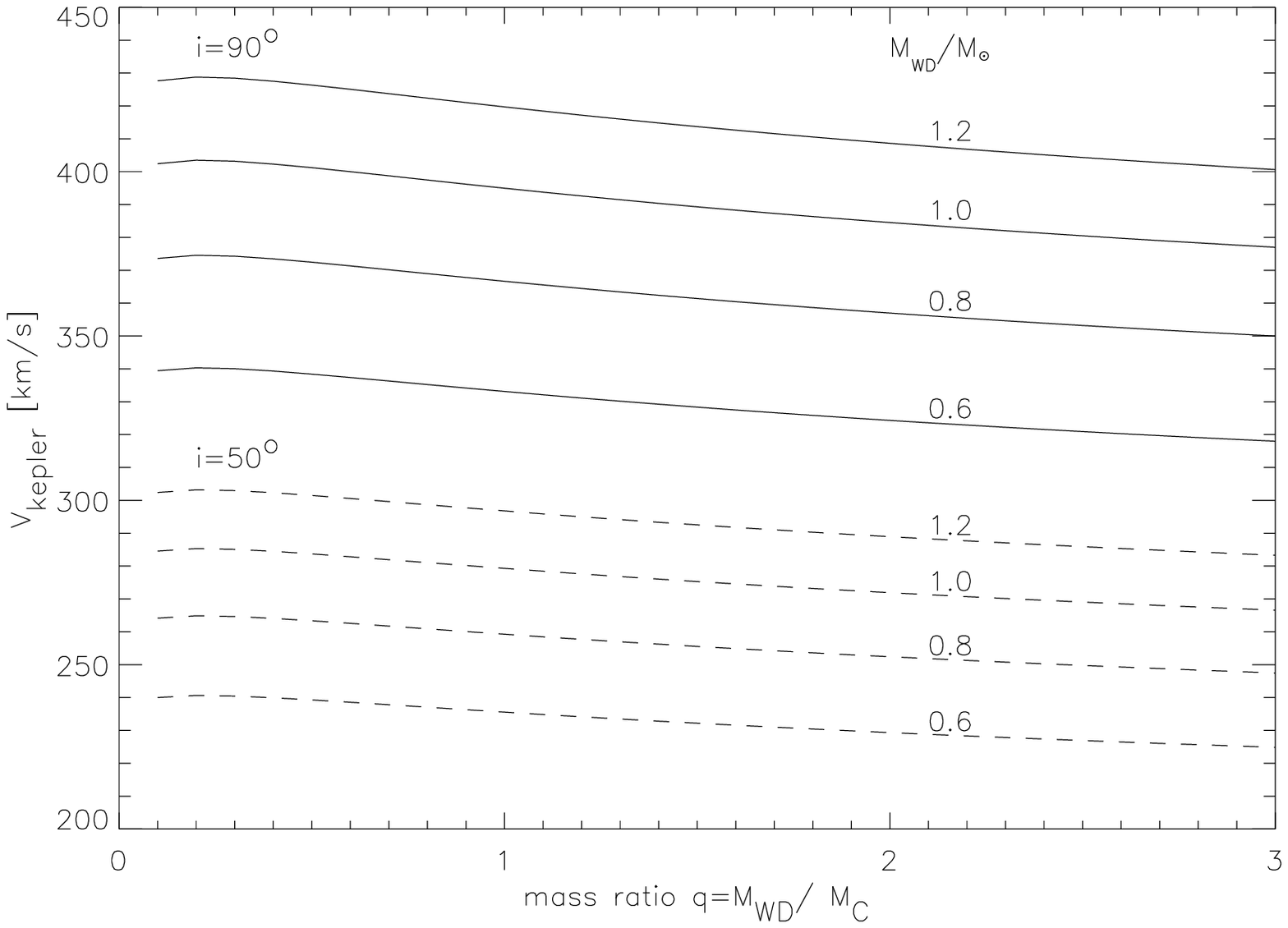}
\caption{Kepler velocity at the rim of the accretion disk at $0.8R_L$
($R_L$ is the volume radius of the Roche lobe of the white dwarf)
over mass ratio $q$ for various masses $M_{\rm WD}$ of the white dwarf
and inclinations $i=50^{\circ}$ and $90^{\circ}$. For the observed
velocities it is clear that the emission lines cannot originate within
the accretion disk. For medium inclinations an origin only at the rim
of the disk is possible.}
\label{kepler}
\end{figure}

The Doppler tomograms (Fig. \ref{tomo}) of our data immediately reveal
that the emission line distribution does not resemble a typical
accretion disk. The usual shape of an accretion disk would appear as a
dark extended ring in the inverse grey-scaled Doppler images,
e.g. compare with Doppler maps of IP Peg obtained with the same
technique in Wolf et al. (1998). Our observed projected velocities are
smaller than 350 kms$^{-1}$. For a binary system with parameters
mentioned above and a disk radius of $r=0.8 R_L=10^{10.95}$ cm ($R_L$
is the volume radius of the Roche lobe of the white dwarf) one finds a
Kepler velocity of 384 kms$^{-1}$ at the rim of the disk. As the
Kepler velocity increases rapidly within the accretion disk, the
observed emission cannot originate within the disk if the inclination
is high. For medium inclinations an origin only at the rim of the disk
is possible.  This conclusion is valid for a huge variety of system
parameters (Fig. \ref{kepler}). Low inclinations are excluded (see
discussion in Sect. 3.2).

%%%%%%%%%%%%        Keplervelocity     %%%%%%%%%%%%%%%%%%%%%%%%%%%%%%%
%
%
%%%%%%%%%%%%%%%%%%%%%%%%%%%%%%%%%%%%%%%%%%%%%%%%%%%%%%%%%%%%%%%%%%%%%

The dim ring in the map of H$_{\alpha}$ at a velocity of $\approx 950$
kms$^{-1}$ is due to the emission from the jet lines.  The Balmer
lines originate at locations with almost no or only very low
velocities. All of the observed flux is symmetrically
distributed around the center of mass extending to velocities of
roughly 350 kms$^{-1}$. 

The intense emission maximum is located near the center of mass.
Part of this emission might come from the irradiated secondary, but as
the maximum of the emission is clearly shifted to $V_Y\approx 0$
kms$^{-1}$ an irradiated secondary probably cannot account for all of
the observed flux.  The Doppler map of H$_{\beta}$ looks
similar. Again, the whole flux including the intense maximum is
distributed symmetrically around the center of mass.

The Doppler maps from the Helium lines are slightly
different. Comparable to the Balmer emission the observed flux
originates at locations with very low velocities. The intensity
distribution as a whole is distributed roughly symmetrical around
the center of mass. But the emission maximum of the Helium lines is
not centered at the center of mass. The coordinates of this intense
maximum are shifted to approximately $V_X$\,$\approx$\,--100 kms$^{-1}$ and
$V_Y$\,$\approx$\,--100 kms$^{-1}$ in both maps. This emission maximum does
not coincide with the gas stream trajectory. But due to its position
in velocity space the emission might originate in the elevated and
radially extended accretion disk rim as proposed by Meyer-Hofmeister
et al. (1998). Thus, the Helium emission might partly
be related to the outermost accretion disk rim. The
\ion{He}{ii} (${\lambda}$\,4686) map also indicates emission from regions
near the white dwarf, which supports our assumptions in Sect. 3.3 for
the estimation of the mass of the secondary.

Another indication about where this material might be located is given
by the fact that the emission is roughly centered on the center of
mass and not on the primary as would be expected for a typical
accretion disk.  This material might perhaps orbit the binary system
around the mass center far outside of both Roche lobes. In a very
simplified estimate one finds a distance of $10^{11.9}$ cm for the
above given binary parameters and a supposed velocity of 150
kms$^{-1}$ for the emitting circum-binary material. This is about 3.5
times the distance of the binary separation.

%____________________________________________________________________

\section{Discussion and summary}

We performed high-speed photometric and spectrophotometric
observations as well as high-resolution spectroscopy of the bright
galactic SSS RXJ0019 between 1992 and 1997.  Our detailed optical
studies result in a new determination of the orbital period. We
observed a color variation in $R-I$ symmetrical to the main minimum
with an amplitude of $\Delta m$\,=\,0.15$^m$. The observed lightcurves
change between having a flat appearance (quiet optical state) and 
showing humps and steps with a 1.8\,h period (excited optical
state). RXJ0019 can change between the two states from night to night.
These variations might be due to short-term changes of the mass
accretion rate from the companion star.

Our spectroscopic investigation shows the well known emission line
spectrum of mainly Balmer and Helium II lines. Symmetric emission
lines adjacent to the Balmer lines show the presence of high velocity
outflows (jets) probably originating near the white dwarf. The jet
lines show an orbital Doppler motion comparable to that of the
\ion{He}{ii} lines.  The velocity of the jet is quite low, indicating
a medium to high inclination of the system.

From the radial velocities we calculated the mass function. For the
assumed range of medium to high inclinations we derive a low mass for
the secondary star ($0.3<M_2<0.5 M_{\sun}$).  Low mass companion stars
might be common among the SSS. This might also be an explanation for
the absence of any spectral features of the companion star in the high
resolution spectra.

We also observe P-Cygni profiles in the Balmer lines showing an
orbital modulation. The P-Cygni absorption almost disappears between
$\phi$\,$=$\,0.7 to 0.9. Higher transitions of the Balmer series up to
H$_{12}$ are only detected by their corresponding P-Cygni absorption.
The velocity of the directed wind responsible for the P-Cygni profiles
is $\approx$\,590\,kms$^{-1}$ with the absorption wings extending to
900\,kms$^{-1}$.

Our trailed spectra show substructures of different components. As
these components are very weak or were observed in lines with bad S/N
ratio the locations of these components cannot clearly be located by
the means of Doppler tomography. The shifted intensity maximum in the
Helium maps at $V_X$\,$\approx$\space--100 kms$^{-1}$ and
$V_Y$\,$\approx$\,--100 kms$^{-1}$ might be due to a radially extended
and elevated accretion disk rim, consistent with theoretical models of
SSS.  But we clearly see that most of the line-emitting material is
not located within the accretion disk. Due to the high accretion rate
and viscous processes involved we assume that, almost the whole
accretion disk is optically thick.  Therefore, no emission from Balmer
and Helium lines originating at the inner parts of the disk can be
observed.

We only see line-emitting material with very low velocities. The
emission distribution is mainly symmetrical around the center of
mass. As material cannot be stationary at the center of mass the
emission must originate at regions with very low velocities. Low
velocities can be found far outside of the binary system. Therefore,
we propose that this SSS has a circumbinary cocoon or disk of hydrogen
and helium responsible for the observed emission lines.

%__________________________________________________________________

\begin{acknowledgements}

The authors would like to thank A. Fiedler for software support and
T. Will for his data of RXJ0019. We are grateful to V. Burwitz and
R. Popham for some helpful remarks on an earlier draft of this
paper. We are grateful to E. Meyer-Hofmeister, F. Meyer and H. Ritter
for helpful discussions. BD is especially grateful to H. Spruit for
his assistance at MPA. We also thank the referee K. Matsumoto for
valuable remarks.

\end{acknowledgements}

%__________________________________________________________________

\end{document}